\documentclass[fleqn,usenatbib]{mnras}

\usepackage{newtxtext,newtxmath}
\usepackage{natbib}
\usepackage{multicol}
\usepackage{xcolor}
\usepackage[T1]{fontenc}
\usepackage{ae,aecompl}

\usepackage{graphicx}	% Including figure files
\usepackage{amsmath}	% Advanced maths commands
\usepackage{amssymb}	% Extra maths symbols
\usepackage{etoolbox}
\makeatletter
\patchcmd\@combinedblfloats{\box\@outputbox}{\unvbox\@outputbox}{}{%
  \errmessage{\noexpand\@combinedblfloats could not be patched}%
}%
\makeatother

\title[A profile in FIRE]{A profile in FIRE: resolving the radial distributions of satellite galaxies in the Local Group with simulations}

\author[J. Samuel et al.]{
Jenna Samuel$^{1}$\thanks{E-mail: jsamuel@ucdavis.edu},
Andrew Wetzel$^{1}$,
Erik Tollerud$^{2}$,
Shea Garrison-Kimmel$^{3}$,
\newauthor
Sarah Loebman\thanks{ Hubble Fellow}$^{1}$,
Kareem El-Badry$^{4}$,
Philip F. Hopkins$^{3}$,
Michael Boylan-Kolchin$^{5}$,
\newauthor
Claude-Andr{\'e} Faucher-Gigu{\`e}re$^{6}$,
James S. Bullock$^{7}$,
Samantha Benincasa$^{1}$,
\newauthor
Jeremy Bailin$^{8}$
\\
$^{1}$Department of Physics, University of California, Davis, CA 95616, USA\\
$^{2}$Space Telescope Science Institute, 3700 San Martin Dr, Baltimore, MD 21218, USA\\
$^{3}${TAPIR, Mailcode 350-17, California Institute of Technology, Pasadena, CA 91125, USA}\\
$^{4}${Department of Astronomy and Theoretical Astrophysics Center, University of California Berkeley, Berkeley, CA 94720, USA}\\
$^{5}${Department of Astronomy, The University of Texas at Austin, 2515 Speedway, Stop C1400, Austin, TX 78712, USA}\\
$^{6}${Department of Physics and Astronomy and CIERA, Northwestern University, 2145 Sheridan Road, Evanston, IL 60208, USA}\\
$^{7}${Department of Physics and Astronomy, University of California, Irvine, CA 92697, USA}\\
$^{8}${Department of Physics and Astronomy, University of Alabama, Box 870324, Tuscaloosa, AL 35487-0324, USA}
}

\pubyear{2019}

\begin{document}
\label{firstpage}
\pagerange{\pageref{firstpage}--\pageref{lastpage}}
\maketitle

\begin{abstract}
While many tensions between Local Group (LG) satellite galaxies and $\Lambda$CDM cosmology have been alleviated through recent cosmological simulations, the spatial distribution of satellites remains an important test of physical models and physical versus numerical disruption in simulations.
Using the FIRE-2 cosmological zoom-in baryonic simulations, we examine the radial distributions of satellites with M$_*>10^5\,$M$_{\odot}$ around 8 isolated Milky Way- (MW) mass host galaxies and 4 hosts in LG-like pairs.
We demonstrate that these simulations resolve the survival and physical destruction of satellites with M$_*\gtrsim10^5\,$M$_{\odot}$.
The simulations broadly agree with LG observations, spanning the radial profiles around the MW and M31.
This agreement does not depend strongly on satellite mass, even at distances $\lesssim$100~kpc. 
Host-to-host variation dominates the scatter in satellite counts within 300~kpc of the hosts, while time variation dominates scatter within 50~kpc.
More massive host galaxies within our sample have fewer satellites at small distances, likely because of enhanced tidal destruction of satellites via the baryonic disks of host galaxies. 
Furthermore, we quantify and provide fits to the tidal depletion of subhalos in baryonic relative to dark matter-only simulations as a function of distance.
Our simulated profiles imply observational incompleteness in the LG even at M$_*\gtrsim10^5\,$M$_{\odot}$: we predict 2-10 such satellites to be discovered around the MW and possibly 6-9 around M31.
To provide cosmological context, we compare our results with the radial profiles of satellites around MW analogs in the SAGA survey, finding that our simulations are broadly consistent with most SAGA systems.
\end{abstract}

% Don't make up new ones. (But that does sound like fun.)
\begin{keywords}
galaxies: dwarf -- galaxies: Local Group -- galaxies: formation -- methods: numerical
\end{keywords}

%%%%%%%%%%%%%%%%%%%%%%%%%%%%%%%%%%%%%%%%%%%%%%%%%%

%%%%%%%%%%%%%%%%% BODY OF PAPER %%%%%%%%%%%%%%%%%%

\section{Introduction}\label{intro}

Dark matter dominates the matter content of dwarf galaxies by up to several orders of magnitude, making them ideal sites for small-scale tests of the standard paradigm for structure formation: cold dark matter (CDM) with a cosmological constant ($\Lambda$). 
CDM makes testable predictions for both the central mass profile of dwarf galaxies and their number density and spatial distribution around more massive host galaxies. 
However, on such small scales, tests of CDM require highly resolved observations that are only feasible within the nearby Universe. 
Fortunately, the Milky Way (MW) and Andromeda (M31) galaxies that make up the Local Group (LG) are host to populations of satellite dwarf galaxies which can provide quantitative tests of CDM on small scales.

LG satellite galaxies have been a source of significant tensions within the CDM model, largely stemming from comparisons of observations to dark matter-only (DMO) simulations that lack the effects of baryonic physics.
Arguably the most famous of these tensions, the ``missing satellites'' problem, describes a discrepancy between the number of subhalos predicted by DMO simulations compared to the smaller number of luminous satellite galaxies observed around the MW \citep[e.g.][]{Moore1999,Klypin1999}.
However, newer simulations that include the effects of baryonic physics through hydrodynamics and sub-grid models show agreement with the number of observed satellite dwarf galaxies in the LG, in part from enhanced tidal disruption of satellites by the baryonic disks of host galaxies \citep[e.g.][]{Brooks2013,Sawala2016,Wetzel2016,GK2018,Kelley2018,Simpson2018,Buck2019}. 
N-body simulations in conjunction with semi-analytic models of galaxy formation have also yielded similar results, showing similar radial distributions for observable satellites \citep[e.g.][]{Maccio2010,Font2011}.
Simultaneously, a better understanding of observational incompleteness has also been critical in alleviating the missing satellites tension \citep{Tollerud2008,Walsh2009,Hargis2014,Kim2018}.

In addition to over-predicting the number of satellites, DMO simulations predict too many dense, massive (``too-big-to-fail") satellites \citep{BK2011,BK2012}, and satellites with steeper (``cuspier") central density profiles than seen in observations \citep{Navarro1996}.
Again, baryonic effects in simulations are a pathway to reconciling these problems because stellar feedback acts to redistribute the central dark matter and ``core-out" the density profile of dwarfs \citep[e.g.][]{Mashchenko2008,Chan2015,Onorbe2015,ElBadry2016,Dutton2016,GK2017,Fitts2017}.

Baryonic effects are also crucial for understanding the predicted phase space coordinates of satellites around simulated MW/M31-like galaxies.
%The LG is the only system where we can measure the full 3D positions and velocities of satellite galaxies and push to resolve the faintest known galaxies.
This phase space information can be used to infer the formation history of satellites and rigorously test CDM predictions.
For example, orbit modeling of the Large Magellanic Cloud (LMC) has provided evidence that it is undergoing its first pericentric passage around the MW, and this may partly be why it is still able to form stars \citep{Besla2007,Kallivayalil2013}.
%Gaia and Hubble Space Telescope (HST) can also resolve the motions of individual stars in the LG, making it possible to measure proper motions of satellites \citep[e.g.][]{GaiaHelmi2018,Kallivayalil2018,Simon2018} and detect diffuse galaxies using the dynamical coherence of their stars \citep[e.g.][]{Ibata2018,Torrealba2018}.
The phase space distribution of LG satellites have further challenged the CDM model because the MW's satellite galaxies appear to be arranged in a thin, planar structure that is coherently rotating, and a similar structure has been found around M31 \citep{LyndenBell1976,Metz2007,Conn2013,Ibata2013,Pawlowski2018}.

Furthermore, the Satellites Around Galactic Analogs (SAGA) survey is broadening our understanding of LG satellites by targeting satellites of MW analogs within 20-40 Mpc of the LG \citep{Geha2017}. 
Their goal is to obtain a complete census of satellites around 100 MW analogs, down to the luminosity of the Leo I dwarf galaxy (M$_r < −12.3$; M$_* \approx5\times10^6$ M$_{\odot}$).
This will make it possible to connect LG satellite galaxies with a large sample of satellite populations, providing a statistically robust cosmological context to interpret LG galaxy formation and evolution.

Satellite dwarf galaxies can be used to study the effects of environment on galaxy formation as well.
Even before satellites accrete onto their host, they are preprocessed by interactions with other dwarf galaxies that are bound to them in small groups \citep[e.g.][]{Zabludoff1998,McGee2009,Wetzel2013,Hou2014}.
Once they fall into their host halo, satellites can be tidally disrupted into diffuse stellar structures by their host.
For instance, the Sagittarius dwarf galaxy is being actively torn apart into a stellar stream within the MW's halo \citep[e.g.][]{LyndenBell1995,Belokurov2006}.
As satellites orbit in the halos of their host galaxies, they are thought to be ram pressure-stripped of their gas, causing their star formation to be subsequently suppressed \citep[e.g.][]{Gunn1972,Fillingham2016}.
The MW and M31 may exert some of the strongest observed environmental effects on their satellite populations: most of their satellites are gas-poor and no longer forming stars, making them an interesting case study for environmental effects \citep[e.g.][]{Einasto1974,Mateo1998,Grcevich2009,McConnachie2012,Slater2014,Spekkens2014,Wetzel2015}.

Given the unique ability to measure full 3D positions and velocities of LG satellites, and thus infer their orbital histories, the LG also provides a fertile physical testing ground for numerical evolution and disruption of subhalos in simulations.
Historically, it has been difficult to use simulations to interpret observations of LG satellites because baryonic simulations have only recently begun to produce dwarf galaxies that do not suffer from numerical over-merging.
Simulations of satellites undergoing tidal disruption have revealed that the most critical simulation parameters in dynamically resolving satellites are spatial and mass resolution \citep[e.g.][]{Carlberg1994,vanKampen1995,Moore1996,Klypin1999a,vanKampen2000,Diemand2007,Wetzel2010,vandenBosch2018}.
While current large-volume simulations can offer a sizable sample of satellite galaxies, their dwarf galaxies have limited resolution both in terms of particle mass and gravitational force softening, which curbs their usefulness in tests that require accurate tidal disruption and survival of satellites.

Understanding formation and evolution is contingent on resolving the radial distribution of satellites as a function of distance from their hosts in cosmological simulations.
Higher resolution, `zoom-in' simulations are now providing satellite populations that are sufficiently well-resolved for studying LG satellite populations in detail.
The main questions this paper aims to answer are: 

\begin{itemize}
    \item Do cosmological zoom-in baryonic simulations reproduce the observed radial distributions of satellites around the MW, M31, and MW analogs?
    \item Do the radial profiles reflect physical disruption from the host galaxy and/or numerical disruption inherent in the simulations?
    \item How do radial profiles in hydrodynamic simulations differ from those in DMO simulations?
    \item If the simulations are representative of the LG, how complete are observations of dwarf galaxies out to large distances around the MW and M31?
\end{itemize}

This paper is organized as follows: in Section~\ref{sims} we describe the simulations used and how satellites were selected from them, in Section~\ref{observations} we describe the observational data set used, in Section~\ref{results} we present our results on radial profiles with comparisons 
of the hydrodynamic simulations to both observations and dark matter-only simulations, and a discussion of the conclusions and implications is given in Section~\ref{conclusion}.

\section{Simulations}
\label{sims}

\subsection{FIRE simulation suite}

We use cosmological zoom-in baryonic simulations from the Feedback In Realistic Environments (FIRE) project\footnote{\url{https://fire.northwestern.edu/}}, run with the upgraded FIRE-2 \citep{Hopkins2018} numerical implementations of fluid dynamics, star formation, and stellar feedback. The FIRE-2 simulations use a Lagrangian meshless finite-mass (MFM) hydrodynamics code, \texttt{GIZMO} \citep{Hopkins2015}. The MFM method allows for hydrodynamic gas particle smoothing to adapt based on the density of particles while still conserving mass, energy, and momentum to machine accuracy. Gravitational forces are solved using an improved version of the N-body \texttt{GADGET-3} Tree-PM solver \citep{Springel2005}, and the gravitational force softening of gas particles automatically adapts to match their hydrodynamic smoothing length.

The FIRE-2 simulations invoke realistic gas physics through a metallicity-dependent treatment of radiative heating and cooling over $10-10^{10}$ K, including free-free, photo-ionization and recombination, Compton, photo-electric and dust collisional, cosmic ray, molecular, metal-line, and fine-structure processes, accounting for 11 elements (H, He, C, N, O, Ne, Mg, Si, S, Ca, Fe).
The cosmic UVB background is included using the \citet{FaucherGiguere2009} model, in which HI reionization occurs early ($z_{\rm reion}\sim10$).
The simulations that we use also model sub-grid diffusion of metals via turbulence \citep{Hopkins2016,Su2017,Escala2018}.

Star formation occurs in gas that is self-gravitating, Jeans-unstable, cold (T $<10^4$ K), dense ($n>1000$ cm$^{-3}$), and molecular (following \citealt{Krumholz2011}).
Each star particle represents a single stellar population under the assumption of a Kroupa stellar initial mass function \citep{Kroupa2001}, and we evolve star particles according to standard stellar population models from \texttt{STARBURST99} v7.0 \citep{Leitherer1999}.
The simulations explicitly model several stellar feedback processes including core-collapse and Type Ia supernovae, continuous stellar mass loss, photoionization, photoelectric heating, and radiation pressure.

For all simulations, we generate cosmological zoom-in initial conditions at $z = 99$ using the \texttt{MUSIC} code \citep{Hahn2011}, and we save 600 snapshots from $z = $99 to 0, with typical spacing of $\lesssim$25 Myr.

We use two suites of simulations in this paper.
The first is the Latte suite of individual MW/M31-mass halos introduced in \citet{Wetzel2016}.
Latte consists of 7 hosts with halo masses M$_{\rm 200m} = 1 - 2 \times 10^{12}$ M$_{\odot}$ (where `200m' indicates a measurement relative to 200 times the mean matter density of the Universe), selected from a periodic volume of length 85.5 Mpc.
Gas and star particles have initial masses of 7070 M$_{\odot}$, though because of stellar mass loss, at $z = 0$ a typical star particle has mass $\approx$ 5000 M$_{\odot}$.
Dark matter particles have a mass resolution of m$_{\rm dm} = 3.5 \times 10^4$ M$_{\odot}$.
Dark matter and stars have fixed gravitational softening (comoving at $z > 9$ and physical at $z < 9$): $\epsilon_{\rm dm} = 40$ pc and $\epsilon_{\rm star} = 4$ pc (Plummer equivalent). The minimum gas resolution (inter-element spacing) and softening length reached in each simulation is $\approx 1$ pc.

In this paper we introduce two new hosts into the Latte suite: m12w and m12r.
We select them using the same criteria as the Latte suite: M$_{\rm 200m}(z = 0) = 1 - 2 \times 10^{12}$ M$_\odot$ and no neighboring halos of similar mass ($> 3 \times 10^{11}$ M$_\odot$) within at least 5 R$_{\rm 200m}$, to limit computational cost.
However, for these two halos we impose an additional criterion: each must host an LMC-mass subhalo.
Specifically, within the initial dark-matter-only simulation, we select halos that host (only) one subhalo within the following limits at $z = 0$: maximum circular velocity V$_{\rm circ,max} = 92 \pm 12$ km/s, distance $d = 51 \pm 40$ kpc, radial velocity $v_{\rm rad} = 64 \pm 17$ km/s, tangential velocity $v_{\rm tan} = 314 \pm 60$ km/s.
These criteria are centered on the observed values for the LMC \citep[e.g.,][]{Kallivayalil2013, vanderMarelKallivayalil2014}, though we use a wider selection window than the observational uncertainties to find a sufficient sample in our cosmological volume, which for this sample is a periodic box of length $172$ Mpc with updated cosmology to match \citet{PlanckCollaboration2018}: $h = 0.68$, $\Omega_\Lambda = 0.69$, $\Omega_m = 0.31$, $\Omega_b = 0.048$, $\sigma_8 = 0.82$, $n_{\rm s} = 0.97$.
The zoom-in re-simulations use the same resolution as the existing Latte suite (given the slightly different cosmology, dark-matter particles have slightly higher mass of m$_{\rm dm} = 3.9 \times 10^4$ M$_\odot$).
While we select these halos to have LMC-like subhalos in the pilot dark-matter-only simulation, when we re-run with baryonic physics the details of the satellite orbit (in particular the orbital phase) do change.
m12w's most massive satellite has M$_* = 8 \times 10^8$ M$_\odot$ and at $z = 0$ is at $d = 248$ kpc, having experienced pericentric passage of 78 kpc 2.4 Gyr ago ($z = 0.19$).
m12r's most massive satellite has M$_* = 2.8 \times 10^9$ M$_\odot$ and at $z = 0$ is at $d = 390$ kpc, having experienced pericentric passage of 30 kpc 0.7 Gyr ago at $z = 0.05$.
We will examine the dynamics of these LMC-like passages in upcoming work (Chapman et al., in preparation).

In addition to the Latte suite, we include one additional individual host (m12z), selected to have a slightly lower halo mass at $z = 0$ and simulated at a higher mass resolution of m$_{\rm baryon,ini} = 4200$ M$_{\odot}$ \citep{GK2018}.

We also use the ELVIS on FIRE suite of two simulations, which selected halos to mimic the separation and relative velocity of the MW-M31 pair in the LG \citep{GK2018}. These simulations have $\approx 2 \times$ better mass resolution than the Latte suite: the Romeo \& Juliet simulation has m$_{\rm baryon,ini} = 3500$ M$_{\odot}$ and the Thelma \& Louise simulation has m$_{\rm baryon,ini} = 4000$ M$_{\odot}$.

All simulations assume flat $\Lambda$CDM cosmologies, with slightly different parameters across the full suite: $h = 0.68 - 0.71$, $\Omega_\Lambda = 0.69 - 0.734$, $\Omega_m = 0.266 - 0.31$, $\Omega_b = 0.0455 - 0.048$, $\sigma_8 = 0.801 - 0.82$, and $n_{\rm s} = 0.961 - 0.97$, broadly consistent with \citet{PlanckCollaboration2018}.

\subsection{Halo finder}

We identify dark-matter (sub)halos using the \texttt{ROCKSTAR} 6D halo finder \citep{Behroozi2013a}.
We identify halos according to their radius that encloses 200 times the mean matter density, R$_{\rm 200m}$, and keep those with bound mass fraction $> 0.4$ and at least 30 dark matter particles.
We generate a halo catalog at each of the 600 snapshots for each simulation.
We then construct merger trees using \texttt{CONSISTENT-TREES} \citep{Behroozi2013b}. 
For numerical stability, we generate halo catalogs and merger trees using only dark matter particles.

We then assign star particles to each (sub)halo in post-processing as follows (adapted from the method described in \citealt{Necib2018}).
Given each (sub)halo's radius, R$_{\rm halo}$, and V$_{\rm circ,max}$ as returned by \texttt{ROCKSTAR}, we first identify all star particles whose position is within 0.8 R$_{\rm halo}$ (out to a maximum radius of 30 kpc) and whose velocity is within 2 V$_{\rm circ,max}$ of each (sub)halo's center-of-mass velocity.
We then keep star particles (1) whose positions are within 1.5 R$_{90}$ (the radius that encloses 90 per cent of the mass of member star particles) of both the center-of-mass position of member stars and the dark matter halo center (thus ensuring the galaxy center is coincident with the halo center), and (2) whose velocities are within 2 $\sigma_{\rm vel}$ (the velocity dispersion of member star particles) of the center-of-mass velocity of member stars.
We then iteratively repeat (1) and (2) until M$_{*}$, the sum of the masses of all member star particles, converges to within 1 per cent.
We keep all halos with at least 6 star particles and average stellar density $> 300$ M$_\odot\,$kpc$^{-3}$.

We examined each galaxy in our sample at $z = 0$ by eye and found that this method robustly identifies real galaxies with stable properties across time; in particular, it reliably separates true galaxies from transient alignments between subhalos and stars in the stellar halos of the MW-mass hosts.
All of the subhalos (within 300 kpc of their host) that we analyze are uncontaminated by low-resolution dark matter.
%with $< 2\%$ contamination (by mass) from low-resolution dark matter,

%%% MAIN TABLE %%%
\begin{table*}
	\centering
	\caption{Host galaxy properties and satellite galaxy counts}
	\label{hosttable}
	\begin{tabular}{llllllll}
		\hline
		Name & M$_{200\rm{m}}$ [$10^{12}$ M$_{\odot}$] & M$_{*}$ [$10^{10}$ M$_{\odot}$] & N$_{\rm sat}$($d<$ 50 kpc) & N$_{\rm sat}$($d<$ 100 kpc) & N$_{\rm sat}$($d<$ 300 kpc) \\
		\hline
		MW & $\sim$1.4 & $\sim$5 & 1 $\pm$ 0.5 (50\%) & 6 $\pm$ 0.5 (10\%) & 13 $\pm$ 0 (0\%) \\
        M31 & $\sim$1.6 & $\sim$10 & 2 $\pm$ 0.5 (25\%) & 5 $\pm$ 1.0 (20\%) & 27 $\pm$ 0.5 (<5\%) \\
        \hline
		m12m & 1.6 & 10.0 & 1 $\pm$ 1.0 (100\%) & 7 $\pm$ 3.0 (45\%) & 27 $\pm$ 2.0 (10\%) \\
		m12b & 1.4 & 7.3 & 0 $\pm$ 0.2 (N/A) & 3 $\pm$ 1.0 (35\%) & 11 $\pm$ 0.7 (5\%) \\
		m12f & 1.7 & 6.9 & 0 $\pm$ 0.0 (N/A) & 1 $\pm$ 1.0 (100\%) & 16 $\pm$ 1.0 (5\%) \\
		Thelma & 1.4 & 6.3 & 1 $\pm$ 0.5 (50\%) & 6 $\pm$ 1.2 (20\%) & 17 $\pm$ 1.2 (10\%) \\
		Romeo & 1.3 & 5.9 & 1 $\pm$ 0.7 (70\%) & 4 $\pm$ 0.7 (20\%) & 17 $\pm$ 1.0 (5\%) \\
        m12i & 1.2 & 5.5 & 0 $\pm$ 0.5 (N/A) & 3 $\pm$ 1.2 (40\%) & 12 $\pm$ 0.5 (5\%) \\
        m12c & 1.4 & 5.1 & 1 $\pm$ 0.9 (90\%) & 8 $\pm$ 1.5 (20\%) & 23 $\pm$ 1.0 (5\%) \\
        m12w & 1.1 & 4.8 & 0 $\pm$ 0.5 (N/A) & 5 $\pm$ 0.9 (20\%) & 22 $\pm$ 1.5 (10\%) \\
        Juliet & 1.1 & 3.4 & 1 $\pm$ 0.5 (50\%) & 8 $\pm$ 1.7 (20\%) & 20 $\pm$ 0.5 (5\%) \\
        Louise & 1.2 & 2.3 & 1 $\pm$ 0.5 (50\%) & 8 $\pm$ 1.5 (20\%) & 23 $\pm$ 0.7 (5\%) \\
        m12z & 0.9 & 1.8 & 1 $\pm$ 0.5 (50\%) & 7 $\pm$ 0.7 (10\%) & 17 $\pm$ 0.7 (5\%) \\
        m12r & 1.1 & 1.5 & 1 $\pm$ 1.2 (120\%) & 5 $\pm$ 1.7 (35\%) & 14 $\pm$ 2.2 (15\%) \\
        \hline
        Time variation &  &  & 1 $\pm$ 0.6 (60\%) & 5 $\pm$ 1.3 (25\%) & 17 $\pm$ 1.1 (5\%) \\
        Host-to-host variation &  &  & 1 $\pm$ 0.5 (50\%) & 5 $\pm$ 2.5 (50\%) & 17 $\pm$ 4.7 (30\%) \\
        Total variation across hosts+time &  &  & 1 $\pm$ 1.0 (100\%) & 5 $\pm$ 3.0 (60\%) & 17 $\pm$ 6.0 (35\%) \\
		\hline
	\end{tabular}
    \begin{flushleft}
    \textbf{(1)} Name of the host.
    \textbf{(2)} Host halo mass (M$_{200\rm{m}}$) at $z=0$. The halo mass of the MW is calculated by taking the value of M$_{200\rm{c}}$ from \citet{BlandH2016} and multiplying by the average value of M$_{200\rm{m}}$/M$_{200\rm{c}}$ for the simulations. The halo mass of M31 is calculated similarly, using the value of M31's M$_{200\rm{c}}$ from \citet{VanderMarel2012}.
    \textbf{(3)} Host stellar mass at $z=0$. The stellar mass of the MW is taken from \citet{BlandH2016}, and the stellar mass of M31 is taken from \citet{Sick2015}. Simulated hosts are ordered from greatest to least stellar mass.
    \textbf{(4-6)} Median and scatter in the cumulative number of satellite galaxies with M$_{*} > 10^5$ M$_{\odot}$ at different distances from the host. For the MW and M31, the scatter is the 68 per cent variation from observational uncertainties from Figure~\ref{lg_obs}. For the individual simulated hosts, the scatter is the 68 per cent variation over time from Figure~\ref{radial_subplots}, spanning 1.3 Gyr ($z=0-0.1$ in steps of $z=0.01$). The percentage in parentheses is the scatter normalized to the median number of satellites. 
    \end{flushleft}
\end{table*}
%%% MAIN TABLE %%%

\subsection{Satellite selection}

We refer to the MW- and M31-mass galaxies in our simulations as ``hosts'' and their surrounding populations of dwarf galaxies with M$_* > 10^5$ M$_{\odot}$ within 300 kpc as ``satellites". 
Each of the eight Latte+m12z simulations contains a single isolated host while each of the two ELVIS on FIRE simulations contains two hosts in a LG-like pair, with their own distinct satellite populations. 
This provides a total of 12 host-satellite systems to study and compare to observations. 
Table~\ref{hosttable} summarizes properties of these systems.
Host galaxies have stellar masses M$_{*}\sim10^{10-11}$ M$_{\odot}$ and dark matter halos M$_{\rm h}=0.9-1.7 \times10^{12}$ M$_{\odot}$.
Host stellar mass is measured by computing the stellar mass enclosed by a 2D radius in the plane of the host disk and a height above and below the plane that together define a cylinder containing 90 per cent of the total stellar mass within a sphere of radius 30 kpc around the host galaxy.

Our satellite selection of M$_* > 10^5$ M$_{\odot}$ corresponds to a minimum of $\sim$20 star particles and a peak halo mass (throughout their history) of M$_{\rm{peak}} > 8 \times 10^8$ M$_{\odot}$ (or $\gtrsim 2.3\times10^4$ dark matter particles prior to infall).
We expect subhalos that contain satellite galaxies with M$_* \sim 10^5$ M$_{\odot}$ to be both resolved in the simulations \citep{Hopkins2018} and nearly complete in observations \citep[e.g.][]{Koposov2007,Tollerud2008,Walsh2009,Tollerud2014,Martin2016}, so we choose this as our lower mass limit to make reasonable comparisons to the MW and M31 (but see Sections~\ref{MW_inc} and~\ref{M31_inc} for further discussion on potential incompleteness in the LG). 
For this analysis, we consider only the total (3D) radial distance from satellite to host galaxy, leaving a complete study of the full 3D positions and the problem of satellite planes for future work. 
For further details on the stellar masses, velocity dispersions, dynamical masses, and star-formation histories of dwarf galaxies in our simulations, see \citet{GK2018,GK2019}.

%%% LOCAL GROUP DATA FIGURE %%%
\begin{figure}
	\includegraphics[width=\columnwidth]{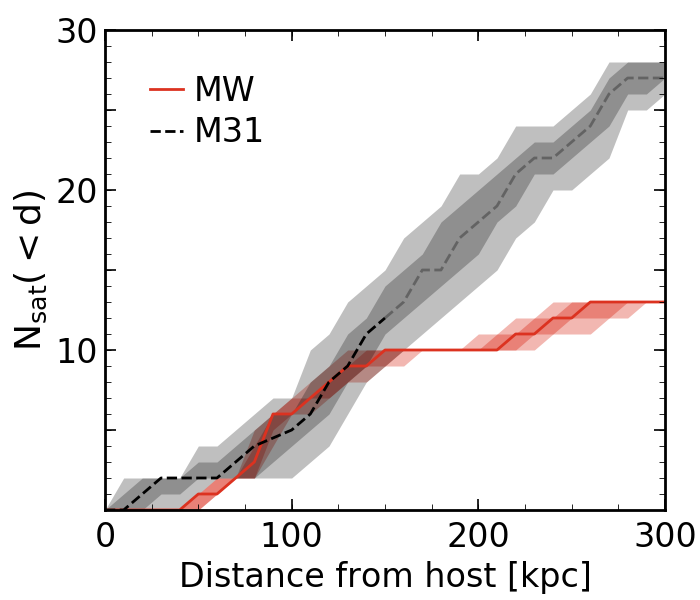}
	\vspace{-7 mm}
    \caption{
    The cumulative number of satellite galaxies with M$_* > 10^5$ M$_{\odot}$ as a function of 3D distance around the MW (red) and M31 (black), similar to Fig. 2 from \citealt{Yniguez2014}.
    M31's line is lighter where the data are known to be incomplete at this stellar mass limit.
    Shaded regions are the 68 per cent and 95 per cent uncertainty in radial distribution when considering the line-of-sight distance uncertainties for satellites.
    Typical 68 per cent (95 per cent) scatter for the MW is $\pm0.3$ ($\pm0.5$) satellites while for M31 it is $\pm1.2$ ($\pm2.4$) satellites.
    The profiles of the MW and M31 are strikingly similar within 150 kpc, but diverge beyond that, where completeness is uncertain.
    We do not attempt to correct the LG observations for completeness.
    }
    \label{lg_obs}
\end{figure}
%%% LOCAL GROUP DATA FIGURE %%%

%%% ALL PROFILES SUBPLOTS FIGURE %%%
\begin{figure*}
	\includegraphics[width=\textwidth]{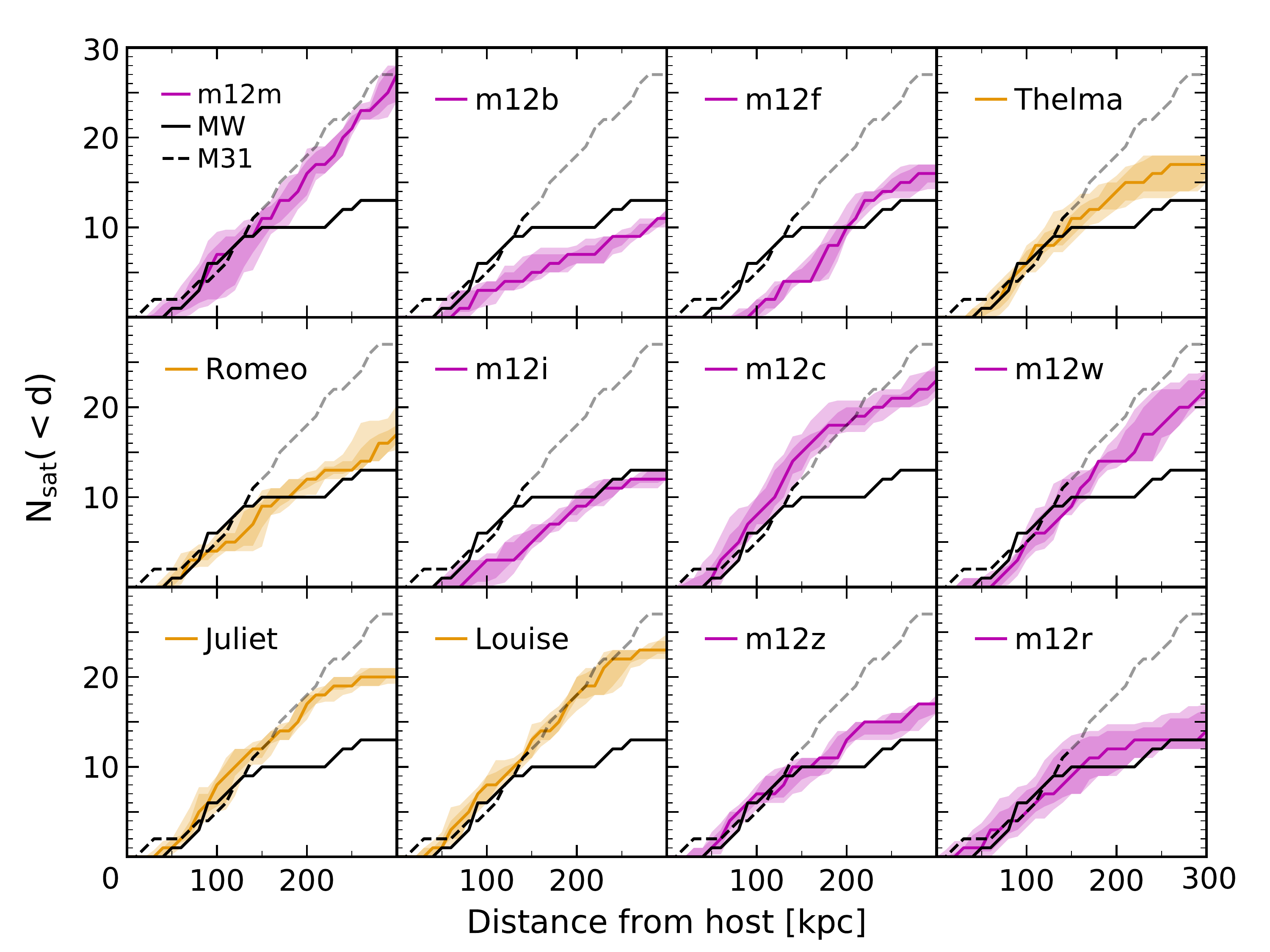}
	\vspace{-7 mm}
    \caption{The cumulative number of satellite galaxies as a function of 3D distance from each host. Results are shown for satellite galaxies with M$_* > 10^5$ M$_{\odot}$.
    Colored lines are the median radial profile of the last 1.3 Gyr ($z=0-0.1$, 11 snapshots in total), and the shaded regions are the 68 per cent and 95 per cent confidence intervals in variation over time.
    Isolated MW-like hosts are pink, while paired LG-like systems are orange.
    Black lines are the median profiles around the MW (solid) and M31 (dashed, lighter where incomplete), taking into account uncertainties in line-of-sight distance to satellites (see Figure ~\ref{lg_obs} for scatter in observed profiles).
    The panels are ordered by decreasing stellar mass of each simulated host galaxy; m12m has the highest M$_* \approx 10^{11}$ M$_{\odot}$ while m12r the has lowest M$_* \approx 1.5 \times 10^{10}$ M$_{\odot}$.
    We do not see any obvious trend in simulated profile shapes or total number of satellite within 300 kpc as a function of host stellar mass.
    Across our sample we find simulated profiles that agree well with both the MW and M31.
    }
    \label{radial_subplots}
\end{figure*}
%%% ALL PROFILES SUBPLOTS FIGURE %%%

\section{Observations}\label{observations}

We use the compilation of observed stellar masses of LG satellite galaxies in \citet{GK2018}, in which they assume stellar mass-to-light ratios from \citet{Woo2008} where available, and elsewhere use M$_*$/L$_{\rm V} = 1.6$ \citep{Martin2008,Bell2001}.
We apply the same stellar mass limit and host-satellite distance limit to the MW and M31 as in our simulation satellite criteria (M$_* > 10^5$ M$_{\odot}$ and d < 300 kpc).
For the satellite galaxies around the MW we take sky coordinates and distances with uncertainties from \citet{McConnachie2012}.
To model the effects of uncertainties in observed distances, we sample MW satellite distances 1000 times assuming Gaussian distributions for the uncertainties to generate a median radial profile with scatter around the MW (Figure~\ref{lg_obs}).

We exclude the Sagittarius dwarf spheroidal galaxy from our MW sample, because it is undergoing significant tidal interactions and we do not believe our halo finder would correctly identify it as a subhalo in the simulations.
We also include two more recently discovered ultra-diffuse satellite dwarf galaxies of the MW: Crater 2 (D$_{\odot} \sim 118$ kpc; \citealt{Torrealba2016}) and Antlia 2 (D$_{\odot} \sim 130$ kpc; \citealt{Torrealba2018}), bringing the total number of MW satellites considered to 13. 
Using the nominal mass-to-light ratio of 1.6 we estimate the stellar masses of these additional galaxies to be M$_* \sim 2.6 \times 10^5$ M$_{\odot}$ for Crater 2 and M$_* \sim 3.4 \times 10^5$ M$_{\odot}$ for Antlia 2.

For the satellite galaxies around M31, we use sky coordinates where available from \citet{McConnachie2012} and apply the same stellar mass and distance restrictions, leaving us with a total of 28 satellite galaxies.
To obtain the 3D radial profiles of M31's satellites with uncertainties, we sample 1000 line-of-sight distances from the posterior distributions published in \citet{Conn2012}, where available.
However, several M31 satellites do not have published distance distributions: M32, NGC205, IC10, And VI, And VII, And XXIX, LGS 3, And XXXI, and And XXXII.
In the cases of M32 and NGC205, they are too close to M31 to reliably determine their distances, so we assume they have the same line-of-sight distance distribution as M31 itself.
Positions on the sky, distances, and distance uncertainties for And XXXI and And XXXII are taken from their discovery paper \citep{Martin2013a}.
For the remaining satellites without line-of-sight distance posteriors, we sample the distances published in \citet{McConnachie2012}, assuming Gaussian distributions on the uncertainties.

Figure~\ref{lg_obs} shows the cumulative number of satellite galaxies around the MW and M31 as a function of 3D distance from the host, and the shaded regions represent estimated scatter in these profiles when we consider uncertainties in line-of-sight distance.
While the sample for M31 includes 28 total satellite galaxies, when we include uncertainties the median number of satellites within 300 kpc is 27.
The resulting 68 per cent scatter averaged across distance from host in LG radial profiles is $\pm$0.3 satellites for the MW and $\pm$1.2 satellites on average for M31. 
We discuss comparisons to the scatter in simulation profiles in Section~\ref{scatter}.

Comparisons to the LG must also be understood in terms of observational completeness.
However, the observational data used for comparison to the simulations in this work have \textit{not} been completeness-corrected.
From the Pan-Andromeda Archaeological Survey \citep[PAndAS,][]{McConnachie2009}, the satellite population around M31 is complete to within 150 kpc (projected) of M31 down to half-light luminosities L$_{1/2} > 10^5$ L$_{\odot}$ \citep{Tollerud2012}.
This includes our lowest satellite galaxy stellar mass limit ($10^5$ M$_{\odot}$), so we think we are making a fair comparison to M31 at least within 150 kpc (where we find evidence for tidal disruption of galaxies by the host, see Sections~\ref{host_mass} and~\ref{dmo_section}).
However, if we assume that our simulations are representative of the LG we find that there may be more galaxies to discover around M31 beyond 150 kpc (see Section~\ref{M31_inc}).
Given that M31 already has a somewhat high number of satellite galaxies compared to the MW, this could potentially make M31's satellite population larger than those of the simulations used here.

Completeness around the MW is complicated by varied survey coverage and seeing through the Galactic disk \citep{Kim2018}.
However, these sources of incompleteness are likely to affect only satellite galaxies fainter than classical dwarf galaxies and therefore they are unlikely to significantly change the results of this work.
Of some concern is the proper identification of diffuse or low surface brightness galaxies (especially through the disk), but this is already being addressed using Gaia data to identify dynamically coherent stellar structures (like the Antlia 2 galaxy included in this work).
We cannot preclude the possibility of further observational incompleteness down to our lowest stellar mass cut out to 300 kpc around the MW and M31.
For this reason, we present comparisons at multiple (higher) stellar mass limits (Sections~\ref{profiles} and~\ref{sat_mass}) and make predictions for the numbers of satellites to potentially be discovered around the MW and M31 (Sections~\ref{MW_inc} and~\ref{M31_inc}).

We also compare our simulations and observations of the LG to results from the Satellites Around Galactic Analogs (SAGA) survey \citep{Geha2017}. SAGA targets MW analogs down to the luminosity of the Leo I dwarf galaxy (M$_r < −12.3$; M$_* \approx5\times10^6$ M$_{\odot}$), and the initial results include the 2D radial profiles of satellite galaxies around 8 MW analogs within 20-40 Mpc of the LG. For more details on how we made this comparison, see Section~\ref{saga_section}.

%%% MAIN COADDED PROFILE FIGURE %%%
\begin{figure*}
    \includegraphics[width=\textwidth]{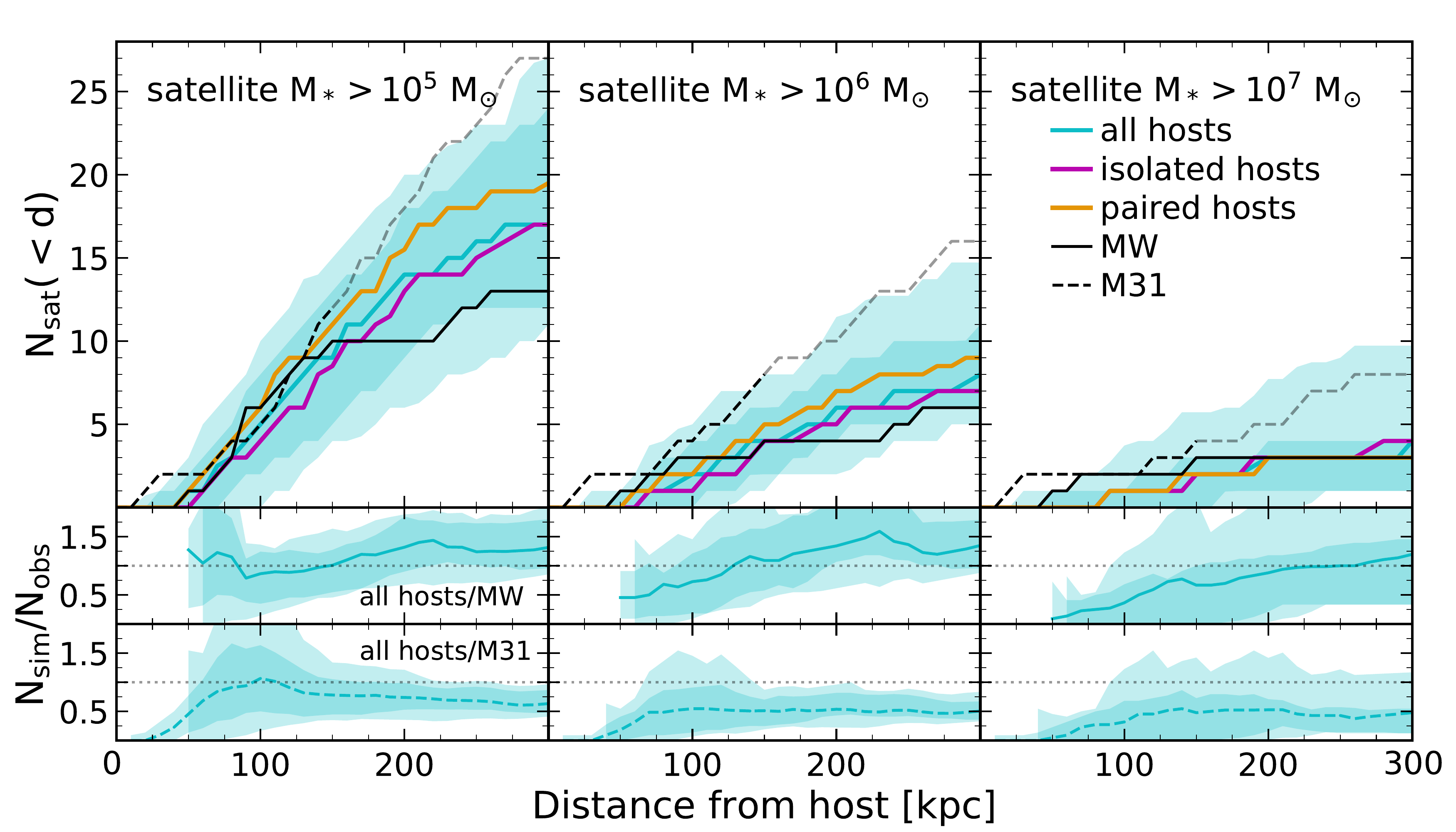}
    \vspace{-5 mm}
    \caption{
    \textit{Top row}: The cumulative number of satellite galaxies across all the simulations and snapshots as a function of 3D distance from the host, for satellites with M$_*$ $> 10^5$ M$_{\odot}$ (left), M$_*$ $>10^6$ M$_{\odot}$ (middle), and M$_*$ $>10^7$ M$_{\odot}$ (right).
    Solid colored lines are the simulation median radial profiles over the last 1.3 Gyr ($z=0-0.1$, using 11 snapshots), while the shaded regions show the 68 per cent and 95 per cent variation.
    We consider all simulations (blue), only the isolated hosts (pink), and only the LG-like paired hosts (orange).
    Black lines are the median radial profiles around the MW and M31, taking into account uncertainties in line-of-sight distance measurements.
    For the two lowest mass bins, the paired hosts have slightly more satellites on average, though this is within the 68 per cent scatter.
    The variation in simulation profiles spans the profiles of the MW and M31 for all three satellite stellar mass bins.
    \textit{Bottom rows}: The median and scatter for all hosts' radial profiles normalized to the observational data for the MW (middle) and M31 (bottom).
    The simulation-to-MW ratio agrees with unity within the 68 per cent scatter at nearly all distances and for all satellite stellar mass limits.
    The simulation-to-M31 ratio agrees with or is close to unity within the 95 per cent scatter at most distances ($\gtrsim$50 kpc) for all satellite stellar mass limits.}
    \label{totalradialdist}
\end{figure*}
%%% MAIN COADDED PROFILE FIGURE %%%

%%% SAGA FIGURE %%%
\begin{figure*}
    \begin{multicols}{2}
	\includegraphics[width=\columnwidth]{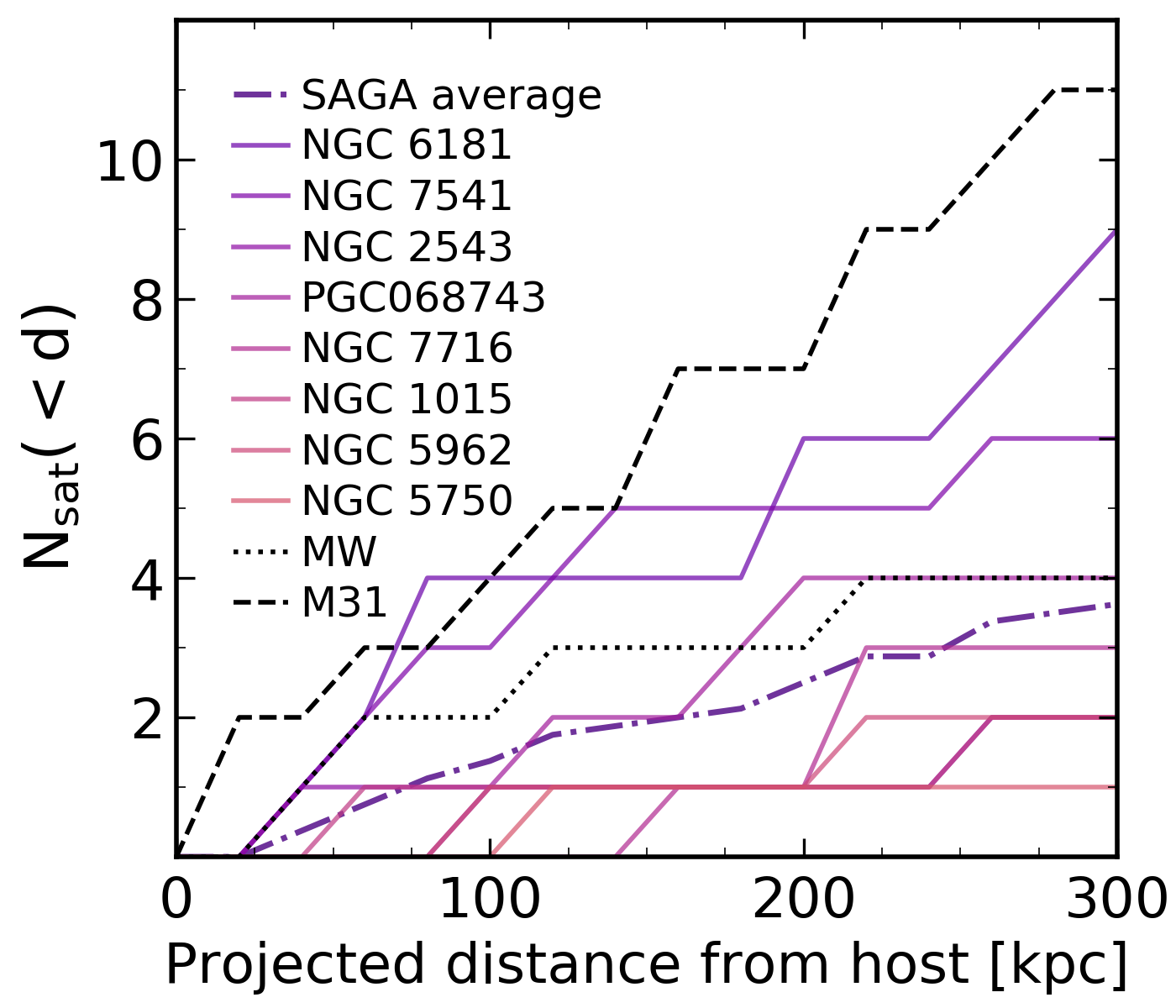}\par
	\includegraphics[width=\columnwidth]{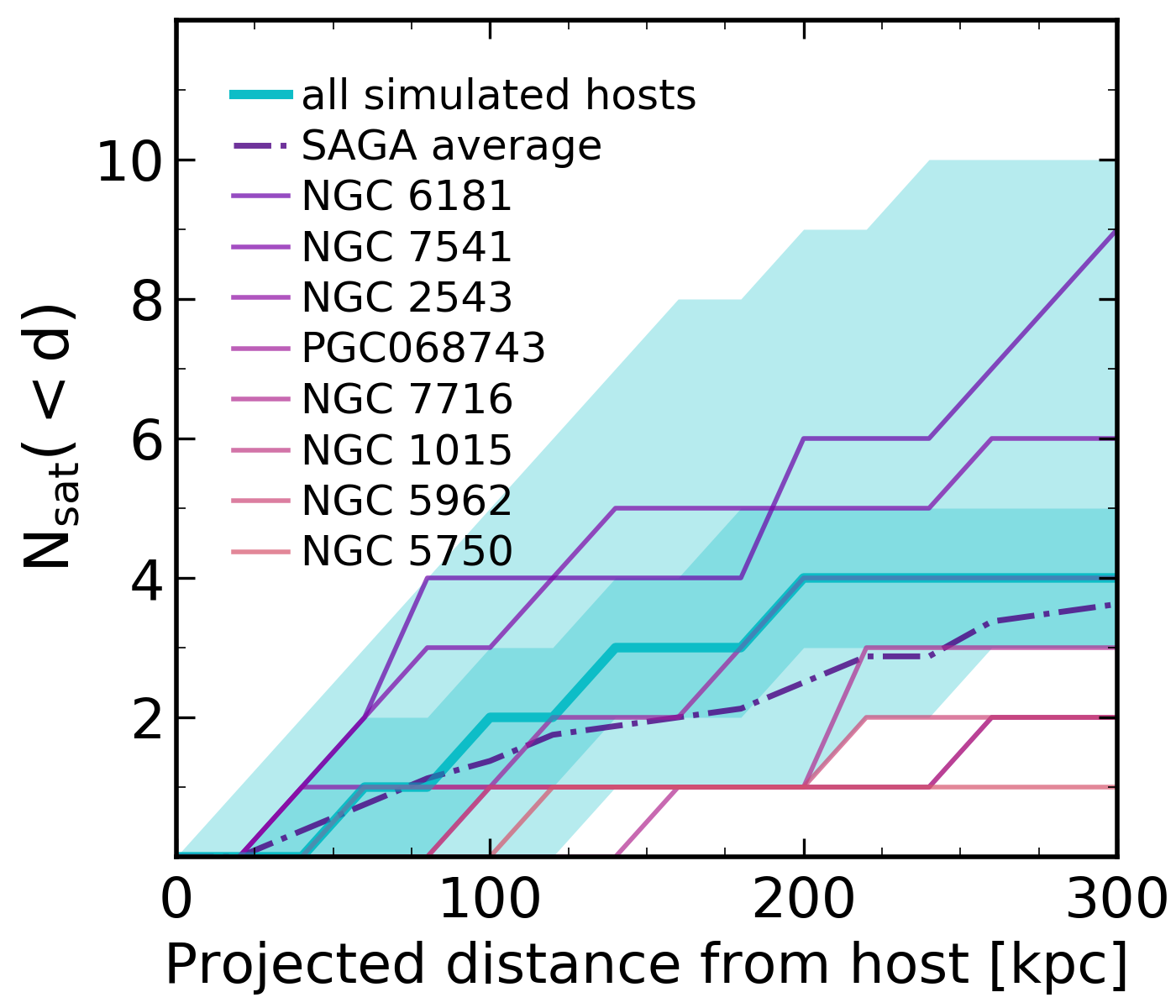}\par
	\end{multicols}
	\vspace{-7 mm}
    \caption{\textit{Left:} Cumulative number of satellite galaxies with M$_{*} > 5 \times 10^6$ M$_{\odot}$ as a function of 2D projected distance, for observations of the LG (black lines) and the 8 complete MW analogs from the SAGA survey and their average (colored lines). 
    The MW lies in the middle of the range of observed profiles, 1-2 satellites above the SAGA average, while M31 is at the upper edge of the distribution of profiles.
    \textit{Right:} Same as left, but showing only SAGA profiles and 2D simulation median profile with scatter (blue). The scatter in the simulations is from random lines of sight, host-to-host variation, and variation over time (but time scatter is not significant).
    Three SAGA hosts have fewer satellites than the 95 per cent simulation limits, but the SAGA average lies mostly within the 68 per cent simulation scatter (and is always within the 95 per cent simulation scatter).
    }
    \label{saga}
\end{figure*}
%%% SAGA FIGURE %%%

\section{Results}\label{results}

We analyze satellite galaxy positions in our simulations over time using halo catalogs from 11 snapshots, taken over $z=0-0.1$ ($\sim$1.3 Gyr) in steps of $\Delta z=0.01$.
We do this for each of the 12 simulated hosts, providing a total of 132 radial distributions of satellite galaxies at different times to study. 
In the inner halo, a typical satellite can undergo a full orbit in under 1 Gyr, while it may take $\sim$3-4 Gyr for a complete orbit in the outer halo. 
This time baseline allows us to time-average over satellite orbits to minimize sampling noise over at least 1/4 of an orbit, which is especially important at small distances where  satellites spend the least amount of time.
Our choice is motivated by a compromise between sampling sufficiently across orbital histories and avoiding systematic redshift evolution (compared with $z = 0$) in the satellite populations.
We find that time-averaging is critical for obtaining accurate results (see Section~\ref{scatter} for results on time variation in radial profiles).

\subsection{Radial profiles}\label{profiles}

Figure~\ref{radial_subplots} shows the cumulative number of satellite galaxies with M$_* > 10^5$ M$_{\odot}$ as a function of 3D distance from the host, or the radial profile, for each individual host-satellite system.
The solid, colored lines are the simulated median radial profile across $z=0-0.1$, and the shaded regions show the 68 per cent and 95 per cent variation over time.
%The time variation in radial profiles is relatively small compared to the total number of satellites for distances $>50$ kpc.
%m12r shows the most scatter over time, but it is still consistent with LG observations.
The median number of satellites within 300 kpc for the simulated hosts ranges from 11-27, consistent with the observed total number of M$_* > 10^5$ M$_{\odot}$ satellite galaxies within 300 kpc of the MW (median N$_{\rm sat}=13$) and M31 (median N$_{\rm sat}=27$) today.
Hosts are ordered by stellar mass with m12m being the most massive (M$_* \approx 10^{11}$ M$_{\odot}$) and m12r the least massive (M$_* \approx 1.5 \times 10^{10}$ M$_{\odot}$).
The number of satellite galaxies does not have an obvious correlation with host mass.
The hosts show a wide range of profile shapes: m12m, m12c, m12w, and Louise closely follow M31, while Thelma, Romeo, m12i, m12z, and m12r more closely follow the MW.

Figure~\ref{totalradialdist} summarizes the key result of this work: the radial profiles of satellite galaxies around the 12 hosts in our simulations span the observed radial distributions of satellites in the LG.
Figure~\ref{totalradialdist} aggregates all of our simulated profiles at three different satellite stellar mass thresholds: M$_*$ $> 10^5$ M$_{\odot}$ (left), M$_*$ $>10^6$ M$_{\odot}$ (middle), and M$_*$ $>10^7$ M$_{\odot}$ (right).
In the top panels we show the median and scatter across all 132 radial profiles simultaneously.
The median radial profile for all simulated hosts (blue) lies on top of the median LG observations at distances $<150$ kpc (where observational completeness is more secure), and at larger distances it lies between the MW and M31 profiles.
The median for paired hosts (orange) is slightly above the total median (blue), while the median for isolated hosts (pink) is slightly below the total median.
However, the paired and isolated medians are still within the total 68 per cent scatter across all the simulations.

The 68 per cent scatter in the simulations overlaps the 68 per cent scatter in MW observations (shown in Figure~\ref{lg_obs}) at nearly all distances, and the MW's median profile is always within the 95 per cent simulation scatter.
M31's median profile lies within the 95 per cent simulation scatter at nearly all distances.
However, M31 appears to have a slight excess of satellites compared to the 68 per cent simulation scatter at small distances ($<$50 kpc) and large distances ($>$250 kpc) for all satellite M$_*$ thresholds, though the uncertainties in M31's profile at small distances are relatively high (see Section~\ref{observations} for more details).
The 95 per cent scatter in simulations overlaps with the 68 per cent scatter in LG profiles at all distances (not shown here, but see Figure~\ref{lg_obs}).
We also a differentially-binned radial distribution for satellites with M$_*$ $> 10^5$ M$_{\odot}$ in Appendix~\ref{diff_radial_appendix}, where we also see general agreement the LG and our simulations.
We conclude that our simulation sample broadly agrees with and spans the profiles around the MW and M31.

In the bottom panels of Figure~\ref{totalradialdist}, we normalize the total simulation median and scatter to the MW and M31 radial profiles.
To calculate the simulated-to-observed ratios, we divide the time-averaged radial profile of each of the 12 simulated hosts by 1000 sampled observational radial profiles of the MW or M31.
Thus, the scatter in each of the bottom panels is from simulated host-to-host variation as well as observational uncertainties.
We find that the MW ratio is consistent with unity at the 68 per cent level at nearly all distances for satellites with M$_*$ $> 10^5$ M$_{\odot}$ and M$_*$ $>10^6$ M$_{\odot}$.
This consistency breaks down at distances $<$150 kpc for M$_*$ $> 10^7$ M$_{\odot}$ given the presence of the LMC and SMC, which are currently near their pericentric passage around the MW.

The M31 ratio is consistent with unity at the 95 per cent level at most distances for satellite M$_*$ $> 10^5$ M$_{\odot}$ and M$_*$ $> 10^7$ M$_{\odot}$, while for M$_*$ $> 10^6$ M$_{\odot}$ the upper scatter in the ratio typically reaches $\sim$0.8.
The M31 ratio is consistent with unity at the 68 per cent level within 50-150 kpc of the host for satellite galaxies with M$_*$ $> 10^5$ M$_{\odot}$.
%The simulations tend to have $<$50 per cent the number of satellites within 50 kpc relative to M31.
Beyond 50 kpc, the median M31 ratio is typically $\sim$50 per cent across the different mass thresholds, indicating that it has a somewhat large satellite galaxy population compared to our average simulation.
This excess of satellite galaxies around M31 relative to the simulations is consistent at all distances, suggesting that M31 may just have more satellites overall, which may mean that its host halo mass is higher than in our simulated sample.
The M31 ratio is most consistent with unity for our lowest mass bin and within 50-100 kpc.
We interpret this, along with our resolution tests in Appendix~\ref{lowres_appendix}, as evidence that we are resolving our sample well even at these lower satellite masses.

Finally, to statistically test whether our simulations' radial profiles are consistent with the LG, we perform a two sample Kolmogorov-Smirnov (KS) test between the median profiles of the LG and each simulated host's profiles (at all 11 snapshots) for satellite galaxies with M$_* > 10^5$ M$_{\odot}$.
The KS test compares the overall shape of the radial profile, and is less sensitive to the absolute number of satellites than taking a ratio between simulations and observations.
We calculate the KS statistic for each of the 11 snapshots over $z=0-0.1$, and we quote the percentage of snapshots where a simulation was \textit{inconsistent} with either the MW or M31.
The KS test results show that a few of the simulations are \textit{inconsistent} with being drawn from the same distribution as the MW at a significance level of 95 per cent: m12f (83 per cent), m12m (27 per cent), m12i (18 per cent), and m12w (9 per cent).
Only m12r (9 per cent) is inconsistent with M31's distribution, and only at one of the 11 snapshots.
We also use the Anderson-Darling (AD) test to check these results and maximize sensitivity to the tails of the radial distributions. 
With the AD tests, we achieve essentially the same results as the KS tests.
We also repeat the KS and AD tests for satellite galaxies with M$_* > 10^6$ M$_{\odot}$, and found that none of the simulated profiles are inconsistent with the MW or M31 at the 95 per cent level, possibly indicating even better agreement at higher masses and that simulations and observations are well resolved and complete in this mass range.

\subsection{Scatter across hosts versus across time}\label{scatter}

Table~\ref{hosttable} summarizes host galaxy mass, number of satellites per host within representative distances, and the scatter over time in each host's radial profile.
We quantify the scatter in radial profiles using the 68 per cent scatter around the median number of satellites with M$_* > 10^5$ M$_{\odot}$ within a given distance from their host.
To understand the importance of time versus host-to-host scatter, we compare the radial profile scatter within individual hosts over time (the pink and orange shaded regions from Figure~\ref{radial_subplots}), scatter among hosts after their time dependence has been averaged out (the solid, colored median lines in Figure~\ref{radial_subplots}), and total scatter among all hosts and snapshots simultaneously (the blue shaded region of Figure~\ref{totalradialdist}).
We quote the 68 per cent scatter about the median in absolute number of satellites and also quote scatter as a percentage relative to the median to give an idea of the fractional variation.
We consider the scatter at three different distances (50, 100, and 300 kpc) to measure time dependence over the full range of the radial profiles.

First, we consider the scatter in the total number of satellite galaxies within 300 kpc.
The combined scatter across all hosts and snapshots within 300 kpc is $\pm$6 satellites, or a 35 per cent variation when normalized to the median of 17 satellites.
The host-to-host scatter after averaging time dependence out is $\pm$4.7 satellites (30 per cent),
whereas the average scatter over time for an individual host is much lower at $\pm$1.1 satellites (5 per cent).
Thus we find that total scatter at large distances is dominated by to host-to-host variations rather than time dependence.

Within 100 kpc, the combined scatter across hosts and time is $\pm$3 satellites (60 per cent), while the host-to-host scatter is $\pm$2.5 satellites (50 per cent), and the time scatter is $\pm$1.3 satellites (25 per cent).
The increased fractional significance of time scatter is likely caused by the relatively small amount of time that satellites spend near pericenter of their orbits.
Within 50 kpc we see that this effect is exacerbated: the combined scatter across hosts and time is $\pm$1 satellite (100 per cent), while host-to-host scatter is $\pm$0.5 satellites (50 per cent), and time scatter is $\pm$0.6 satellites (60 per cent).
We conclude that at large distances ($\gtrsim$100 kpc) the total scatter across all 132 radial profiles is dominated by host-to-host variation, and at small distances ($\lesssim$50 kpc) the total scatter is dominated by time dependence from satellite orbits.
%However, we note that our time baseline of 1.3 Gyr may not be sufficient to sample full orbits of the most distant satellites which could mean we are missing some time variation in radial profiles at larger distances.

\subsection{Comparison to the SAGA survey}\label{saga_section}

We also compare our simulated and LG profiles to the on-the-sky projected radial profiles of 8 MW analogs in the SAGA survey.
To match the SAGA luminosity limit, we select satellite galaxies for comparison to SAGA in our simulations, the MW, and M31 by requiring them to have stellar masses above the value of Leo I, M$_* \approx 5 \times 10^6$ M$_{\odot}$. 
We generate 2D projections of the simulations along 1000 lines of sight for each of the 12 host-satellite systems at 11 snapshots, from which we compute the median and scatter across the simulated sample. 
For M31 satellites, we use only their projected on-the-sky distances from M31, assuming a line-of-sight distance to M31 of 780 kpc.
For the MW, we use the 3D positions of the satellites and their line-of-sight distance uncertainties to generate 2D realizations from 1000 lines of sight as we did for the simulations.

Figure~\ref{saga} (left) shows the observed 2D profiles for SAGA hosts, the MW, and M31. 
Most SAGA systems have fewer satellite galaxies compared to the MW and M31, which could be an effect of the broad mass selection function used in SAGA to choose MW analogs within uncertainties on the MW's stellar mass \citep{Geha2017}.
Because our simulations show only slightly higher satellite counts in our LG pairs compared with isolated hosts (see Figure~\ref{totalradialdist}), this implies that the SAGA selection of \textit{isolated} hosts is unlikely to be a significant cause of difference as compared with the LG.
The MW profile lies in the middle of the SAGA sample, and its scatter via line-of-sight averaging spans most of the range between the average SAGA profile and M31's profile within 200 kpc.
M31 still has a relatively large number of satellites compared to the SAGA observations at all distances (especially beyond 150 kpc), but two of the SAGA hosts have numbers of satellites approaching M31's profile.

Figure~\ref{saga} (right) shows the SAGA profiles compared to the simulations.
The blue line is the median and the shaded regions show the 68 per cent and 95 per cent scatter in simulations.
The scatter in simulated 2D profiles is mainly due to host-to-host variation and line-of-sight averaging, while time variation contributes a negligible amount of scatter in projection.
At distances $>$100 kpc, three of the eight SAGA hosts are at or below the 95 per cent simulation limits.
The SAGA average lies within the 68 per cent simulation scatter at most distances, though still slightly below the simulation median for distances $>$100 kpc.
The best agreement between the SAGA average and the simulations is at small distances ($<$100 kpc), where they overlap the most.
Overall, the simulation scatter encompasses five of the eight SAGA profiles and we find reasonable agreement among SAGA results, the LG, and our simulations in projection.

%%% SATELLITE MASS FIGURE %%%
\begin{figure}
	\includegraphics[width=\columnwidth]{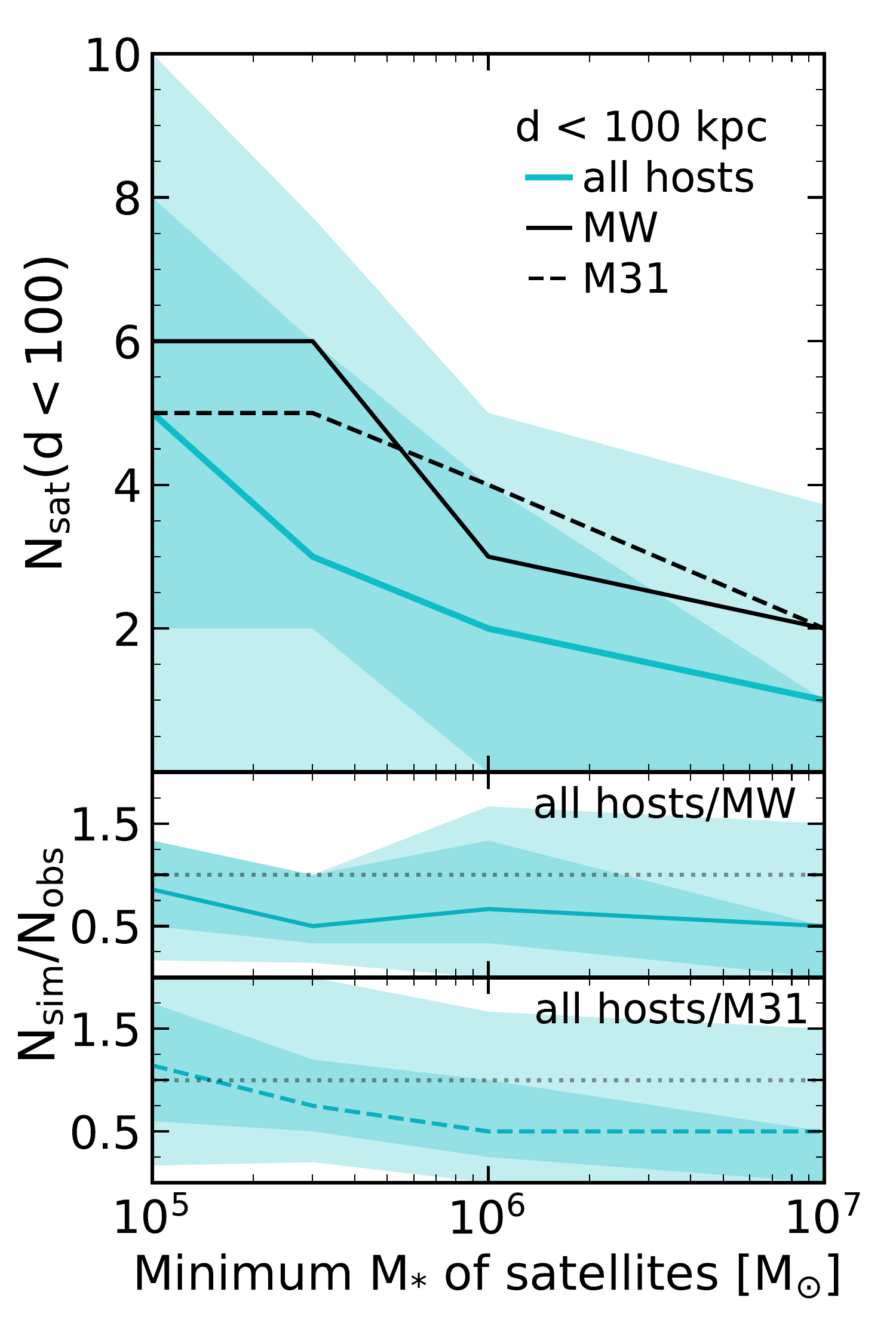}
	\vspace{-7 mm}
    \caption{\textit{Top panel:} The number of satellite galaxies with M$_* > 10^5$ M$_{\odot}$ within 100 kpc of their host as a function of minimum stellar mass. The blue line and shaded regions show the median, 68 per cent, and 95 per cent variation over the last 1.3 Gyr ($z=0-0.1$, using 11 snapshots) across all of the simulations. The simulation median is $\lesssim$2$\times$ smaller than observations of the MW and M31, but the scatter in simulations encompasses the MW and M31 at all satellite masses.
    \textit{Bottom panels:} The median and scatter in the ratio of N$_{\rm sat}$($<$d) in the simulations relative to observations of the MW and M31. The trend in the ratios is essentially flat with increasing minimum satellite stellar mass. Even if the simulations have fewer satellites on average within 100 kpc, less massive satellites (hence closer to the resolution limit) are not preferentially under-represented or over-disrupted in the simulations compared to observations.}
    \label{nsatvsstellarmass}
\end{figure}
%%% SATELLITE MASS FIGURE %%%

%%% HOST MASS FIGURE %%%
\begin{figure*}
    \begin{multicols}{2}
	\includegraphics[width=\columnwidth]{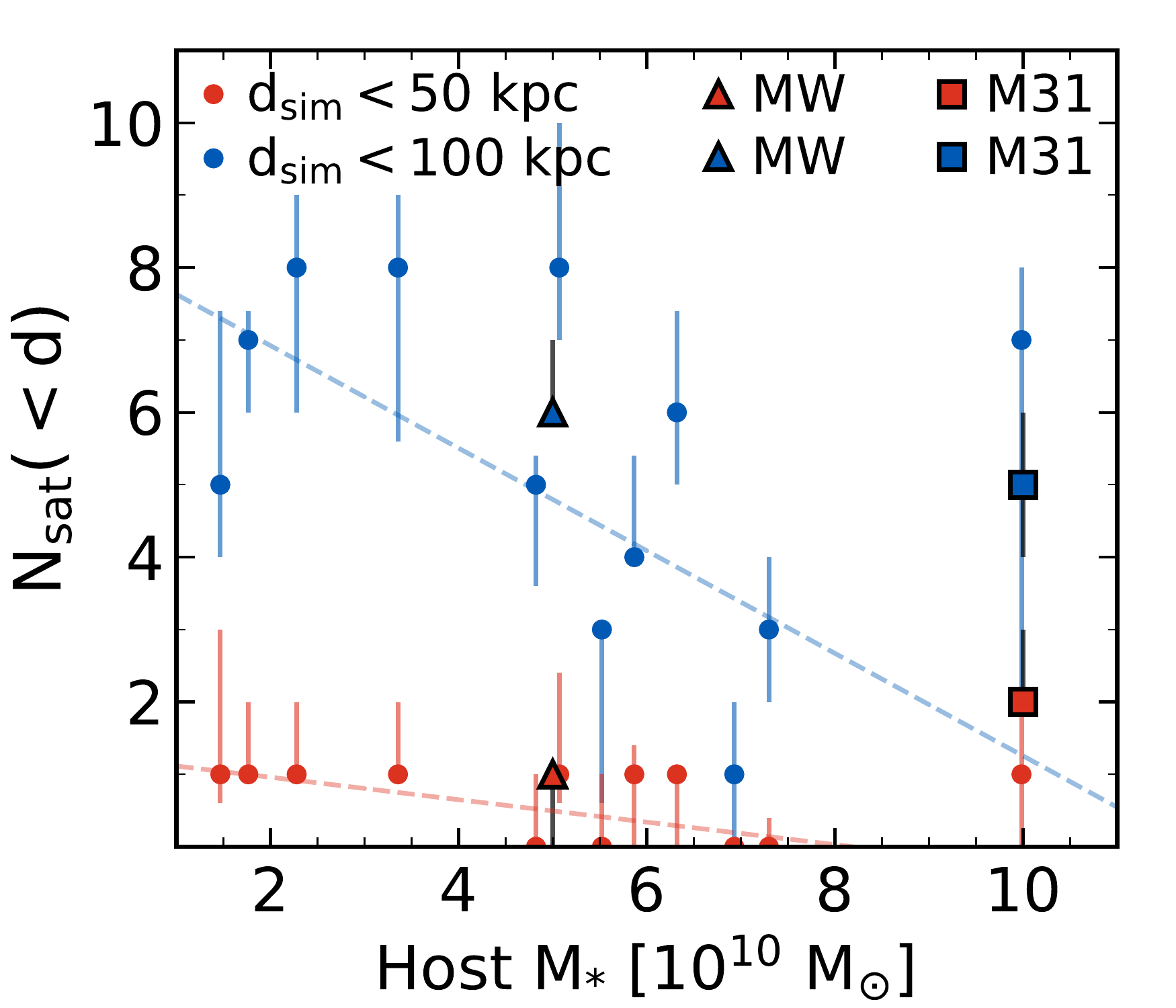}\par
	\includegraphics[width=\columnwidth]{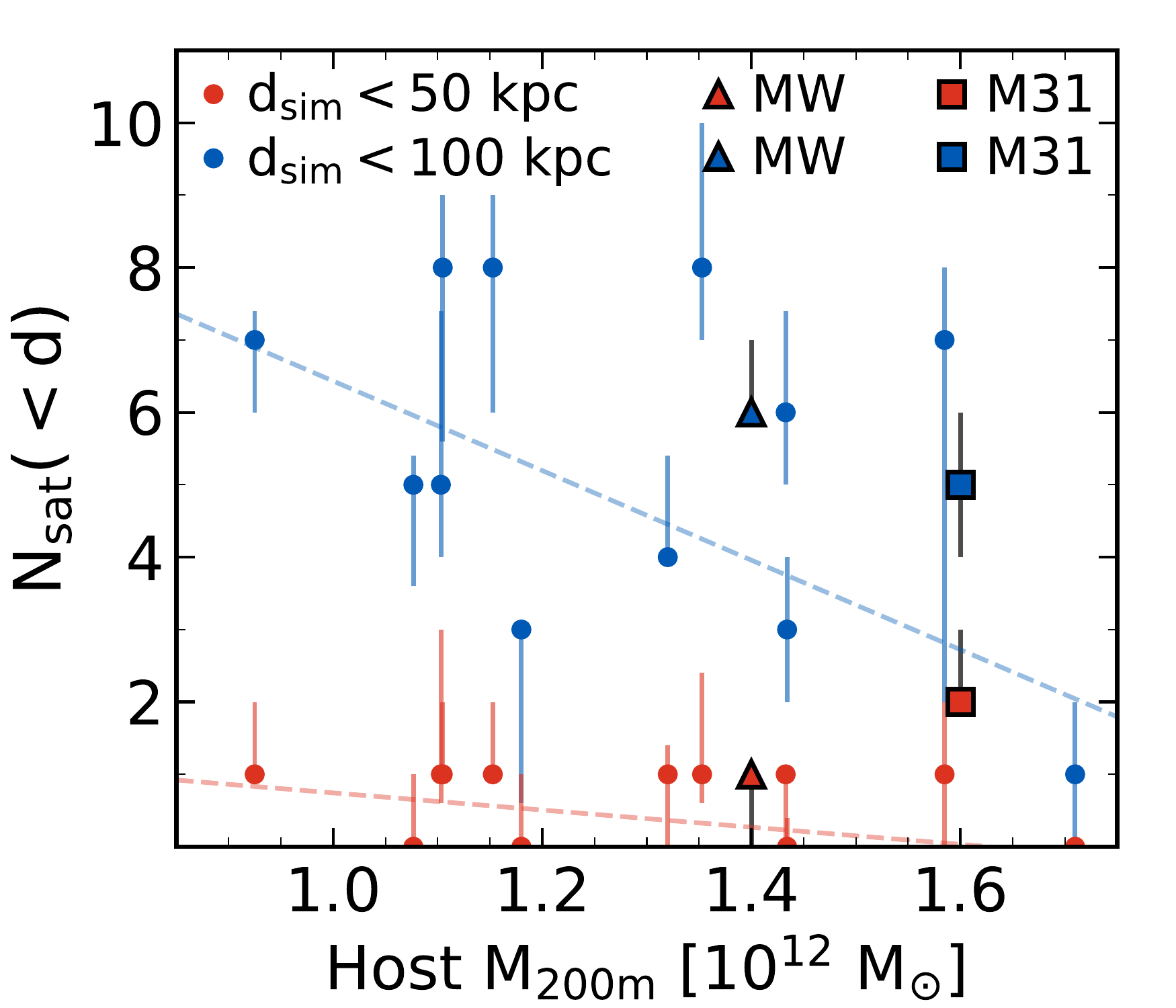}\par
	\end{multicols}
	\vspace{-7 mm}
    \caption{The number of satellite galaxies (M$_* > 10^5$ M$_{\odot}$) within 50 (red) and 100 (blue) kpc of each host as a function of host stellar mass (left) and host halo mass (right). Circles with error bars are the simulated host medians and 68 per cent variation over the last 1.3 Gyr ($z=0-0.1$, using 11 snapshots). Observations of the MW are triangles and observations of M31 are squares, and their error bars are from uncertainties in line-of-sight distances. We use a linear fit to the simulations to demonstrate the negative trends. 
    \textit{Left:} The number satellites decreases with increasing host stellar mass within both 50 and 100 kpc of the host. The red points show that there are little to no satellites within 50 kpc in the simulations. 
    The MW has a number of satellites comparable to the simulations, and M31 is within simulation variation at the high mass end.
    \textit{Right:} Same as left, but using the halo mass (M$_{200\rm{m}}$) of the host. There are similar trends in the number of satellite galaxies nearby their host, but the number of satellite galaxies within 50 kpc is less correlated with halo mass than it is with stellar mass. 
    Though not shown, the number of satellite galaxies within 300 kpc is essentially uncorrelated with host stellar and halo mass (see Fig. 3 of \citealt{GK2018} for satellite counts within 300 kpc as a function of host halo virial mass).
    This indicates that the host's stellar mass is a better predictor of the survival of satellite galaxies within 50 kpc.
    }
    \label{nsat_v_host_mass}
\end{figure*}
%%% HOST MASS FIGURE %%%

\subsection{Dependence on satellite galaxy stellar mass}\label{sat_mass}

In this section, we examine in more detail whether our results within small distance ($\lesssim$100 kpc) depend on the stellar mass of satellite galaxies.
This is a test of how our simulations compare to observations across our satellite mass range, and because higher-mass satellites are better resolved in both stellar mass and halo mass, it is also a test of dependence on resolution.
Here, we assume that satellites with larger stellar masses inhabit more massive dark matter halos, but this may not always be true given scatter in the galaxy stellar mass-halo mass relation \citep{GK2017a,Fattahi2018}.
We use the number of satellite galaxies within small distances as our summary statistic because this is where we expect to see the most prominent effects of tidal disruption and perhaps numerical over-merging in simulations. 
However, given the small numbers of satellites within 50 kpc of the hosts in both our simulations and the observations, we choose 100 kpc as the limiting distance as a reasonable trade-off between testing at small distances and obtaining reasonable statistics.

The top panel of Figure~\ref{nsatvsstellarmass} shows the median number of simulated satellite galaxies within 100 kpc of their host across all hosts and snapshots as a function of the lower limit on satellite stellar mass, compared to the MW and M31.
We consider satellite stellar mass limits from M$_* > 10^5$ M$_{\odot}$ up to M$_* > 10^7$ M$_{\odot}$, the highest stellar mass bin where we still have sufficient statistics in our simulations.
The observed medians for the MW and M31 are $\lesssim$2$\times$ higher than the simulation median.
This difference for satellites with M$_* > 10^7$ M$_{\odot}$ could be caused by the presence of the LMC and SMC near their pericenters around the MW, which is not typical in a time-averaged sense.
Even so, the 95 per cent simulation scatter always encompasses the observations, and the 68 per cent scatter mostly contains the MW and M31 lines.

In the bottom panels, we normalize our simulations to the MW and M31 observations, by sampling from both the simulated hosts and observational uncertainties simultaneously.
%If we were near our resolution limit, we would expect increasingly fewer satellites relative to observations at lower masses, where the satellites are less well resolved.
In general, satellite galaxies with smaller stellar masses reside in less massive dark matter halos, so they are resolved with fewer star and dark matter particles.
Therefore, in the absence of confounding numerical artifacts, we might expect our simulations to show increasingly fewer satellites relative to observations at lower stellar masses if we are reaching our resolution limit.
Interestingly, we find the best agreement with observations when we include our lowest mass satellite galaxies (M$_* > 10^5$ M$_{\odot}$).
The simulated-to-observed ratios are always consistent with unity at the 95 per cent level, but are only consistent with unity at the 68 per cent level when we include satellites with M$_* > 10^{5-6}$ M$_{\odot}$.
The trend in the ratios as a function of minimum satellite stellar mass considered is relatively flat, though our simulations may not be producing as many higher-mass satellites as the LG.
This is broadly consistent with results from \citet{GK2018}, who examined all satellites out to 300 kpc and found that most hosts are consistent with the MW and M31's satellite population is only slightly larger than the simulations.
Therefore, relative to observations, our simulations do not suffer from obvious over-destruction of satellites at the stellar masses that we consider.

To more thoroughly analyze numerical resolution, we test for convergence of the radial distributions of subhalos in these simulations compared to those from lower resolution simulations in Appendix~\ref{lowres_appendix}.
There, we show that our (high resolution) simulations are converged to within $\sim20$ per cent on average, and the 68 per cent (host-to-host) scatter is consistent with 100 per cent convergence at distances $>30-40$ kpc.
We note that this exercise suffers from the effects of an additional disruptive effect in the low-resolution simulations: because the low-resolution host galaxies have $\sim 2 \times$ larger stellar masses, their subhalos may be more easily tidally stripped or destroyed as they orbit close to the host.
This may have the effect of making the high-resolution simulations appear less converged, at least at small distances from the host.
We conclude that subhalos hosting the satellite galaxies in our high-resolution simulations are sufficiently resolved for tests of the satellite populations' radial distributions.
For a more nuanced discussion of convergence and additional resolution tests, see Appendix~\ref{lowres_appendix}.

%Therefore, the simulations do not appear to suffer from over-destruction of satellites at the low mass end of our sample (see also Appendix~\ref{lowres_appendix}) compared to observations, but our simulations may not be producing as many higher-mass satellites as the LG.

\subsection{Dependence on host mass}\label{host_mass}

We test whether our results for satellite galaxies with M$_* > 10^5$ M$_{\odot}$ are sensitive to the stellar and halo masses of the host galaxies within 50 and 100 kpc.
Figure~\ref{nsat_v_host_mass} (left) shows the median number of satellite galaxies within 50 and 100 kpc of the host as a function of host stellar mass. 
The simulations agree well with the MW, and while the simulation trends lie below M31, the scatter for the most massive simulated host (m12m) is still consistent with M31.
Within both 50 and 100 kpc of the host there is a negative trend in the number of satellites as a function of host galaxy stellar mass, and 4 hosts have no satellites at all (median over time) within 50 kpc. 
These 4 hosts all have stellar masses $\gtrsim 5 \times 10^{10}$ M$_{\odot}$, which is the average host stellar mass for the simulations. 
We interpret this and the trend lines as evidence for enhanced tidal destruction of satellites in our simulations due to the increased gravitational potential from the more massive hosts' baryonic disks.

Figure~\ref{nsat_v_host_mass} (right) shows the median number of satellites within 50 and 100 kpc as a function of host halo mass (M$_{200\rm{m}}$). 
When controlling for host halo mass, the time variation or scatter in the simulations is consistent with the MW and M31.
However, M31 lies above both the simulation trends and the MW lies slightly above the 100 kpc trend line.
M31 on the other hand, lies above the simulation trends, but still within the simulation scatter. 
The trend in N$_{\rm sat}$(d<100 kpc) as a function of host halo mass is slightly less steep than as a function of host stellar mass for the simulations.
The correlation of N$_{\rm sat}$(d<50 kpc) with host mass is stronger for stellar mass (Pearson correlation coefficient: $r_*=-0.32$) than it is for halo mass ($r_{200\rm{m}}=-0.22$).
The correlations of N$_{\rm sat}$(d<100 kpc) with each type of host mass are: $r_*=-0.35$ and $r_{200\rm{m}}=-0.43$.
Within 300 kpc (not shown) we find little to no correlation: $r_*=0.14$ and $r_{200\rm{m}}=0.07$.
We interpret the larger correlation with host stellar mass within 50 kpc and steeper trend with host stellar mass within 100 kpc as the destructive tidal effects of the host baryonic disk manifesting at sufficiently small distances. 
Since host stellar mass is more correlated with satellite count within 50 kpc, we conclude that host stellar mass is a better predictor of the total number of surviving satellite galaxies within 50 kpc of the host, where we expect disk effects to be strongest.

Naively, we might expect the number of satellites at any distance to correlate positively with halo mass, and because M$_*$ correlates with M$_{200\rm{m}}$, we might also expect a similar correlation with stellar mass. 
Both the negative trend with host stellar mass and the lack of satellites around the more massive galactic disks suggest instead that the baryonic disk is depleting the satellite population at small distances.
However, we note that because of the correlation between host M$_*$ and M$_{200\rm{m}}$ in our simulations (see Figure~\ref{hostmasscolor}), we cannot strictly disentangle the tidal effects of the separate disk and halo components of the host independently in our analysis.
Despite this uncertainty, we find it physically plausible that tidal destruction of satellites can negate our initial expectations, at least for satellites closer to the host galaxy, consistent with results presented in \citet{GK2017} and \citet{Kelley2018} that show a lack of satellites or subhalos at small distances in the presence of a disk potential.
This also explains the lack of correlation between the number of satellites within 300 kpc and host halo mass (also noted in Fig 3 of \citealt{GK2018} as a function of host halo virial mass): while increasing halo mass increases the number of expected satellites, the correlation of host stellar mass with host halo mass and the tidal destruction from the host disk act to cancel out this dependence, at least within the limited host mass range that we explore with our simulations.

We also note that while our simulated hosts have a wide range of stellar masses (M$_{*}\sim10^{10-11}$ M$_{\odot}$), they were selected over only a narrow range in host halo mass (M$_{200\rm{m}}\sim1-2\times10^{12}$ M$_{\odot}$).
Therefore, our sample is missing MW/M31-like host galaxies with much larger (or smaller) halo masses, but with stellar masses that scatter into our sample's range.
Hosts with more extreme halo masses like this could potentially lead to a less negative correlation of N$_{\rm sat}$ with host M$_*$ within 100 kpc.

%%% DMO FIGURE %%%
\begin{figure}
	\includegraphics[width=\columnwidth]{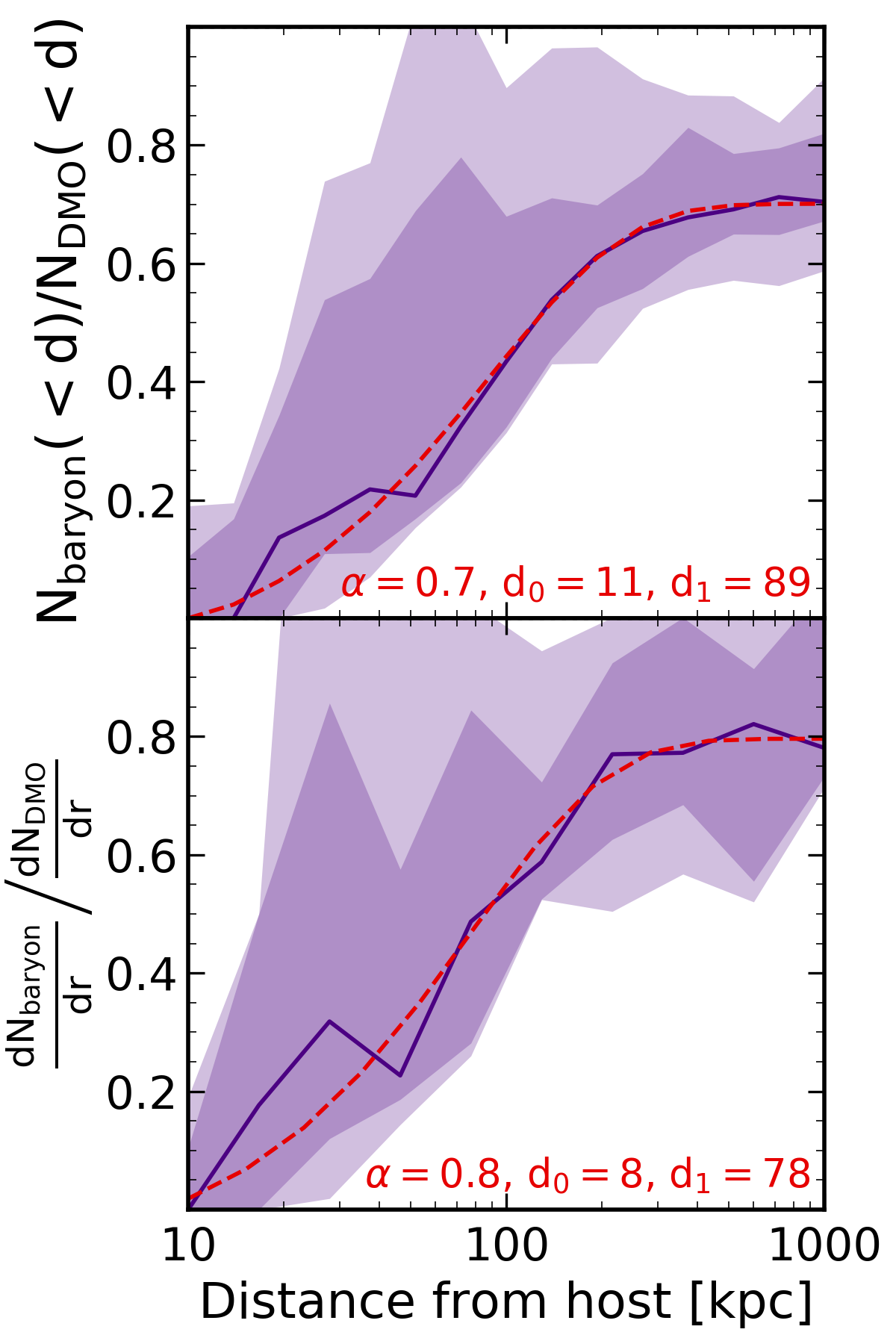}
	\vspace{-7 mm}
    \caption{The ratio of the number of subhalos, at a given subhalo mass, in baryonic versus dark matter-only (DMO) simulations. Purple line and shaded regions are the host-to-host median and scatter in the baryonic-to-DMO ratio as a function of 3D distance. Red line is an analytic fit to the ratio, and fit parameters are also shown in red. Subhalos were selected to have M$_{\rm peak} > 8 \times 10^8$ M$_{\odot}$, to mimic the halo masses of dwarf galaxies in the baryonic runs. \textit{Top panel:} Baryonic-to-DMO ratio for cumulative subhalo counts as a function of distance. Relative to the DMO simulations, the baryonic simulations have $\sim$70 per cent (median) the number of subhalos beyond 200 kpc. The ratio drops rapidly within this distance, where the DMO subhalos are not subject to the additional gravitational potential of a host's baryonic disk. The median ratio declines to zero within $\sim$15 kpc of the host.
    \textit{Bottom panel:} Same as above, but showing differential subhalo counts (discrete distance bins). The ratio is $\sim$80 per cent (median) beyond 200 kpc, and it declines to zero within $\sim$10 kpc of the host.
    }
    \label{dmo}
\end{figure}
%%% DMO FIGURE %%%

\subsection{Comparison with dark matter-only simulations}
\label{dmo_section}

The seven Latte simulations also have dark matter-only (DMO) versions run with the same number of DM particles and the same force softening\footnote{However, the DMO simulations have DM particles with slightly higher masses of m$_{\rm dm}=4.2\times10^4$ M$_{\odot}$ due to the lack of baryons. We correct for this by multiplying DMO subhalo masses by $1-f_{\rm b}$ to account for the mass that would be otherwise relegated to baryons given the cosmic baryon fraction ($f_{\rm b} \equiv \Omega_{\rm b}/\Omega_{\rm m}$) of our baryonic simulations.}.
We compare the DMO versions to the baryonic simulations in order to investigate the effects of baryonic physics on the radial profiles of subhalos. 
To compare with the baryonic simulations, we find that satellite galaxies with M$_* > 10^5$ M$_{\odot}$ have typical peak dark matter halo masses M$_{\rm peak}\gtrsim8\times10^8$ M$_{\odot}$.

We select subhalos in the DMO and baryonic simulations by requiring them to be within 1000 kpc of their host and to have M$_{\rm peak}>8\times10^8$ M$_{\odot}$ so their halos are approximately as well-resolved as baryonic satellites down to M$_* \sim 10^5$ M$_{\odot}$.
We then average the radial profiles of each host-subhalo system over $z=0-0.1$ using all available snapshots (67 total) for improved subhalo statistics at small distances.
We compute the ratio of a host's baryonic-to-DMO profiles for each host individually, and then examine the median and scatter across hosts.
We compute the ratio as a function of distance for both cumulative and differential subhalo counts: N($<d$) and N($d_1<d<=d_2$), respectively.
Figure~\ref{dmo} shows the cumulative ratio of baryonic-to-DMO subhalos (top) and the differential ratio (bottom).
The line and shaded regions are the median and scatter showing only host-to-host variation, as the time dependence has been averaged out prior to taking the ratio.

Within $\sim$100 kpc from the hosts, there are $\lesssim$50 per cent the number of baryonic subhalos compared to DMO subhalos, and this continues to rapidly drop as distance decreases until the (median) ratio reaches zero at 10-15 kpc.
As \citet{GK2017} studied extensively using embedded disk potentials in DMO simulations of m12i and m12f, this is almost entirely due to the presence of the additional gravitational potential from the disk in the baryonic simulations. 
Here, we provide a more robust sample of simulations where we also time-average for each host, which is critical given how little time satellites spend near pericenter.
The scatter within 100 kpc is greater than the scatter at 200-1000 kpc, due to a few hosts that have a number of baryonic subhalos closer to their number of DMO subhalos at small distances.

At large distances ($\gtrsim$200 kpc), the median ratios of baryonic-to-DMO subhalos flatten to $\sim$0.7 for the cumulative case and $\sim$0.8 for the differential case.
This indicates that baryonic effects can reduce the masses of halos even at large distances from the host by $\sim$20-30 per cent as compared with DMO simulations.
The overall reduction of substructure in the baryonic simulations relative to the DMO simulations is likely due to a combination of various baryonic effects such as reionization through our UV background and environmental effects like ram-pressure stripping and interactions with large scale structure such as filaments \citep{Benitez2013}.
Any of these processes may act to blow out gas from the galaxies in our baryonic simulations, shallowing their gravitational potential significantly in lower-mass galaxies like dwarfs, which in turn allows for easier removal of dark matter mass through gravitational interactions.
\citet{Sawala2017} also found that the abundance of for subhalos with masses below $10^{9.5}$ M$_{\odot}$ in the \texttt{APOSTLE} simulations was reduced at all distances out to 200 kpc from the hosts.
For the largest distances they examine, between 50 and 200 kpc, \citet{Sawala2017} found a reduction in substructure abundance of 23 per cent.
This is similar to our results for the ratio of baryonic-to-DMO differential profiles between about 200 to 1000 kpc, where where we see a reduction in substructure abundance of about 20 per cent.

We provide fits to the ratio of baryonic-to-DMO subhalo counts as a function of distance that may be used to estimate the number of subhalos containing satellite galaxies (M$_* > 10^5$ M$_{\odot}$) in other DMO simulations.
We fit to the median ratio across hosts, and use the 68 per cent variation in the ratio as uncertainty on the fitted median values\footnote{The $z=0$ snapshots of baryonic m12i, m12f, and m12m are publicly available at \url{ananke.hub.yt} for comparison to individual hosts.}.
In Table~\ref{dmotable}, we also explore fits using other subhalo mass cuts.
We fit the median of the cumulative and differential baryonic-to-DMO ratios as a function of distance ($d$):

\begin{equation}
    f\left(d\right) =
    \begin{cases} 
      0 & 0 \leq d < d_0 \\
      \alpha\left[1-e^{-\frac{d-d_0}{d_1}}\right] & d \geq d_0 
   \end{cases}
\end{equation}\label{dmo_equation}

Where $\alpha$ is the asymptotic value of the ratio for infinitely large $d$, $d_0$ is the inner cutoff where the ratio goes to zero, and $d_1$ is the distance within which the ratio sharply declines.
For the cumulative profile shown we find: $\alpha=0.7$, $d_0=11$ kpc, and $d_1=89$ kpc.
For the differential profile shown we find: $\alpha=0.8$, $d_0=8$ kpc, and $d_1=78$ kpc.
Table~\ref{dmotable} shows these parameters for other fits using instantaneous bound halo mass for subhalo selection (not shown in Figure~\ref{dmo}).

We find that, as expected, the fitted baryonic-to-DMO subhalo count ratios tend towards close to unity at large distances and drop to zero near the baryonic disk boundary.  
The fits indicate that even at arbitrarily large distances from the host, the baryonic subhalos are subject to additional destructive baryonic effects.
The decline in the fitted ratios within $\sim$100 kpc is strikingly sharp: the cumulative and differential ratios both go to zero within $\sim10$ kpc, indicating the physical boundary of intense gravitational effects from the baryonic disk.
We see the same general trends in fits, for both cumulative and differential ratios, across the three different subhalo selection methods we use.

Our results agree with studies that have found that satellite survival depends on host-satellite distance at pericentric passage \citep[e.g.][]{DOnghia2010,Zhu2016,Sawala2017,GK2018b,Nadler2018,RodriguezWimberly2019}.
We note that the destruction that we see is somewhat less extreme than in \citet{GK2017}, who used two of our baryonic simulations (m12i and m12f) and found no surviving subhalos at $z = 0$ within $\sim$20 kpc of the host. 
Our results here are more robust given the larger host sample and that we time-average the profiles.

\citet{Kelley2018} examined the destructive effects of an analytical disk+bulge potential embedded in DMO simulations, where the analytical potential was allowed to realistically grow over time to match the MW's potential at $z=0$.
They found the ratio of subhalo counts that were subject to the embedded potential relative to subhalo counts that were not subject to the additional potential to be $\sim$1/3 within 50 kpc of their hosts. 
We find that our baryonic simulations are more efficient at destroying subhalos within 50 kpc, with a median ratio of baryonic-to-DMO subhalo counts of $\sim$1/5 at this distance.
This could mean that additional baryonic effects, such as supernovae, in our simulations lead to enhanced modulation of the baryonic-to-DMO ratio. 
However, the simulations used in \citet{Kelley2018} were calibrated to the mass of the MW and may not capture the full effects of our wider mass range which encapsulate more massive M31-like galaxies as well.

\citet{Sawala2017} performed a similar comparison of the radial distributions of substructure in baryonic and DMO simulations, averaging over time and four hosts from the \texttt{APOSTLE} simulations. 
However, the baryonic disks of the hosts in their simulations are $\sim 2 \times 10^{10}$ $M_{\odot}$, which is lower than the average disk masks of our hosts.
Thus, based on our results from Section~\ref{host_mass}, we expect to see more substructure destroyed around our hosts than \citet{Sawala2017} found.
They found a baryonic-to-DMO ratio of subhalo counts of $\sim 1/2$ at $d < 10$ kpc, and $\gtrsim 3/4$ at $d > 50$ kpc. 
By comparison, we see a much smaller median baryonic-to-DMO ratio of zero within $10$ kpc of our hosts, but the host-to-host scatter reaches as high as $\sim 1/5$ at $d < 10$ kpc. 
At $d > 50$ kpc, the scatter in our ratio varies from $\sim 1/5 – 1$ and at $d > 100$ kpc it is $\gtrsim 1/2 -1$. 
\citet{Newton2018} repeated this exercise and found subhalo depletion similar to \citet{Sawala2017}: their baryonic-to-DMO ratio ranged from $\sim 1/2$ at small distances and rose to $\sim 4/5$ at large distances ($R_{200}$) from the host. 
Considering the differences in the stellar masses of the host disks between our simulations, the larger subhalo depletion we see at small distances compared to that from \citet{Sawala2017,Newton2018} is unsurprising, and we note that far from the host disk our results are more similar to each other.

%%% DMO FITS TABLE %%%
\begin{table}
	\centering
	\caption{Parameters for fits to the ratio of subhalos in baryonic versus dark matter-only simulations in Equation 1. Cumulative distributions refer to the total number of subhalos enclosed as a function of 3D distance while differential distributions refer to discrete bins in 3D distance.}
	\label{dmotable}
    \begin{tabular}{llll}
    \hline
    Subhalo selection method & $\alpha$ & $d_0$ [kpc] & $d_1$ [kpc] \\ \hline
    \textit{Cumulative distributions} &  &  &  \\
    M$_{\rm peak} > 8 \times 10^{8}$ M$_{\odot}$ & 0.7 & 11 & 89 \\
    M$_{\rm bound} > 10^{8}$ M$_{\odot}$ & 0.8 & 13 & 106 \\    
    M$_{\rm bound} > 10^{7}$ M$_{\odot}$ & 0.8 & 2 & 98 \\ \hline
    \textit{Differential distributions} &  &  &  \\
    M$_{\rm peak} > 8 \times 10^{8}$ M$_{\odot}$ & 0.8 & 8 & 78 \\
    M$_{\rm bound} > 10^{8}$ M$_{\odot}$ & 0.9 & 21 & 95 \\
    M$_{\rm bound} > 10^{7}$ M$_{\odot}$ & 0.9 & 0 & 100 \\ \hline
    \end{tabular}
    \begin{flushleft}
    %\textbf{Note:}
    \end{flushleft}
\end{table}
%%% DMO FITS TABLE %%%

%%% CONCENTRATION FIGURE %%%
\begin{figure*}
    \begin{multicols}{2}
	\includegraphics[width=\columnwidth]{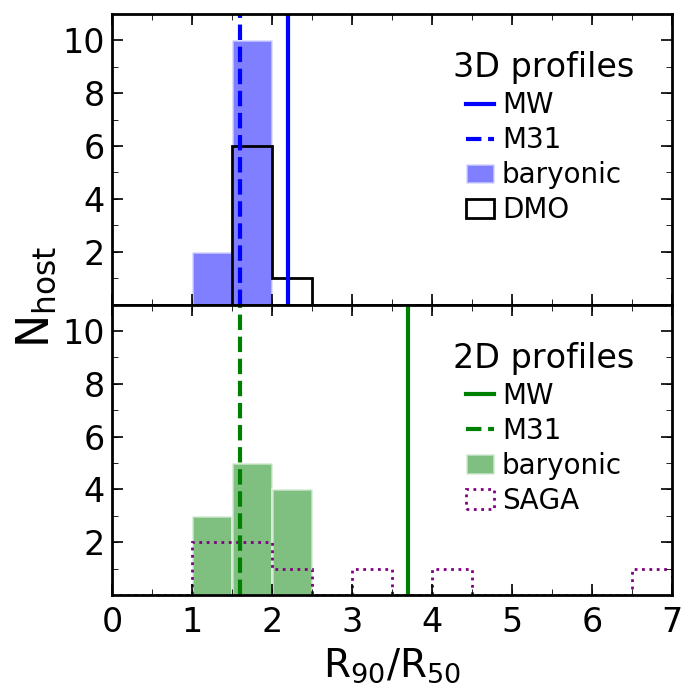}\par
	\includegraphics[width=\columnwidth]{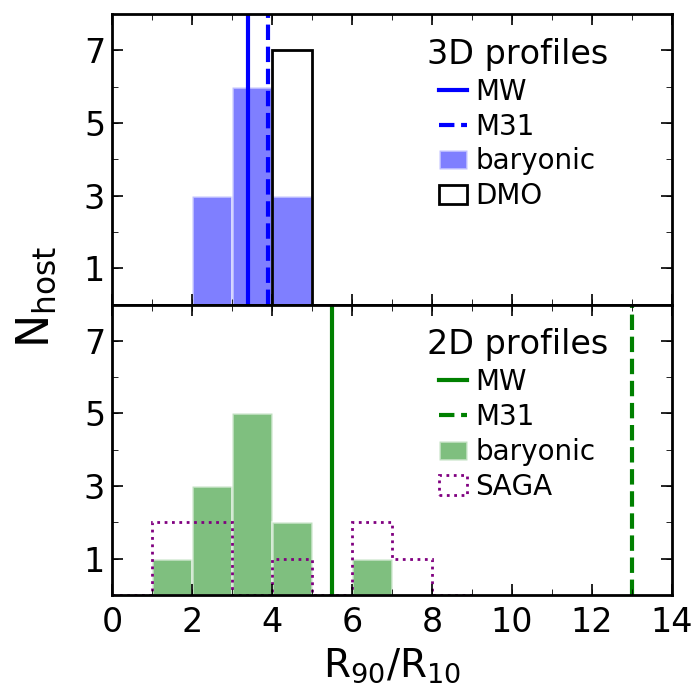}\par
	\end{multicols}
	\vspace{-7 mm}
    \caption{Radial concentration values of the simulated and observed profiles. Top panels are 3D profiles of all baryonic satellites with M$_*>10^5$ M$_{\odot}$ (M$_{\rm peak}>8\times10^8$ M$_{\odot}$ for DMO), and bottom panels are 2D projections of profiles for satellites with M$_*>5\times10^6$ M$_{\odot}$ for comparison to the SAGA survey.  Filled color histograms are baryonic simulations, unfilled black histograms are DMO simulations, colored vertical lines are the MW and M31, and dashed unfilled histograms are SAGA systems. 
    \textit{Left:} Concentration as measured by $R_{90}/R_{50}$. The 3D simulated and observed profiles (top) have a narrow range of concentration values. The MW agrees better with the slightly higher concentrations of the DMO simulations, while M31 agrees with both the baryonic and DMO simulations. The 2D profiles (bottom) of the baryonic simulations, M31, and most of SAGA lie in the same narrow range as the 3D profiles. However, the MW and 3 of the SAGA hosts have much higher concentration.
    \textit{Right:} Same as left, but for $R_{90}/R_{10}$. The 3D profiles are distributed over a narrow range in concentration with DMO simulations tending to have higher concentration. Both the MW and M31 agree with the baryonic simulations. The 2D profiles are spread over a wider range. The MW, 3 SAGA hosts, and one baryonic host have slightly higher concentration, while M31 is much more concentrated than any of the other systems in projection.
    }
    \label{concentration}
\end{figure*}
%%% CONCENTRATION FIGURE %%%

\subsection{Radial concentration}

We further quantify satellite radial profiles using their shape, which we refer to as radial concentration. 
A profile with higher concentration generally has more of its satellites at small distances than at large distances from the host. 
We parameterize the concentration of our simulated and observed radial profiles using two metrics: $R_{90}/R_{50}$, the ratio of the distances enclosing 90 per cent and 50 per cent of the total number of satellite galaxies around a host, and $R_{90}/R_{10}$ to be sensitive to variations in satellite counts at smaller distances.

We analyze the concentration of 3D profiles considering LG satellites, baryonic simulation satellites, and DMO simulation subhalos that are within 300 kpc of their host. 
We measure concentration of the baryonic profiles for satellite galaxies with M$_*>10^5$ M$_{\odot}$ around each of the 12 baryonic hosts and in the LG.
DMO subhalos were selected as in Section~\ref{dmo_section}, by requiring M$_{\rm peak}>8\times10^8$ M$_{\odot}$ for each of the 7 available DMO hosts.
We also analyze the concentration of profiles in 2D projection for LG satellites, simulated baryonic satellites, and SAGA survey satellites with M$_*>5\times10^6$ M$_{\odot}$.
We report the concentration of each simulated host as the median over 11 snapshots from $z=0-0.1$, and the observed LG values as the median across 1000 sampled profiles.

Figure~\ref{concentration} (left) summarizes $R_{90}/R_{50}$ concentration measurements for the baryonic simulations, DMO simulations, the LG, and the SAGA survey.
$R_{90}/R_{50}$ does not significantly differentiate baryonic simulations from DMO simulations.
M31's $R_{90}/R_{50}$ agrees with both the baryonic and DMO simulations, but the MW's $R_{90}/R_{50}$ is slightly higher than the baryonic simulations, and is more consistent with the DMO simulations.
However, we do find that $\sim$10-30 per cent of individual snapshots for half of the baryonic hosts (m12b, m12c, m12r, m12z, Romeo, and Thelma) have $R_{90}/R_{50}$ values that are at least as concentrated as the MW.
This suggests that the MW has a slightly more concentrated profile shape relative to the median values for each baryonic simulation host.
In 2D projection, the MW appears more highly concentrated than the baryonic simulations and M31.
Most of the 2D SAGA profiles over the baryonic simulation distribution, but two SAGA systems have much higher concentration that is closer to the MW and one SAGA system has a concentration nearly twice that of the MW.

Figure~\ref{concentration} (right) shows $R_{90}/R_{10}$ concentration measurements for the simulations and observations.
The baryonic simulations cover a broader range of values for $R_{90}/R_{10}$ than they do for $R_{90}/R_{50}$.
DMO simulations tend to have systematically higher average $R_{90}/R_{10}$ than the baryonic simulations. 
Thus, the primary difference between baryonic and DMO profiles lies in the fraction of satellites at small distances ($\lesssim$100 kpc), where the DMO simulations have a larger fraction of their subhalos.
This is consistent with the results of Section~\ref{dmo_section}, where we show that the largest discrepancies between baryonic and DMO profiles occur within $\lesssim$100 kpc of the hosts.
The $R_{90}/R_{10}$ values for the MW and M31 are consistent with the baryonic simulations and lie outside the range of DMO values.
The 2D profile span an even broader range in $R_{90}/R_{10}$ than the 3D profiles.
The SAGA systems are broadly consistent with the baryonic simulations, with a few more SAGA systems lying at the high end of the baryonic distribution.
The MW in projection is also near the higher end of the baryonic simulations, and M31 appears much more concentrated than anything else.

The concentrations of the MW and M31 profiles generally overlap with the concentrations of the simulated baryonic profiles.
Considering incompleteness in M31's satellite population, if there are more M31 satellites to discover beyond 150 kpc, it could potentially push M31's $R_{90}$ higher.
This could increase M31's concentration to a point where it is discrepant with the baryonic simulations. 
However, when using the $R_{90}/R_{50}$ metric, the MW is slightly more radially concentrated and therefore less consistent with the baryonic simulations than the DMO simulations.
The MW in 2D projection appears more concentrated than most of the baryonic simulations, and M31 in projection is more concentrated than anything else.
The SAGA profiles mostly overlap the projected baryonic simulation profiles, with a few SAGA systems having higher concentration more like the MW.
Our results indicate that the MW may not be as much of a high-concentration outlier as previously thought: \citet{Yniguez2014} noted that the MW had a larger concentration than all of their DMO simulations.
Concentration depends strongly on observational completeness assumptions though, and this may hint that there are more satellites just above M$_* = 10^5$ M$_{\odot}$ remaining to be discovered at farther distances from the MW as we explore next.

%%% MW PREDICTIONS FIGURE %%%
\begin{figure*}
    \begin{multicols}{2}
	\includegraphics[width=\columnwidth]{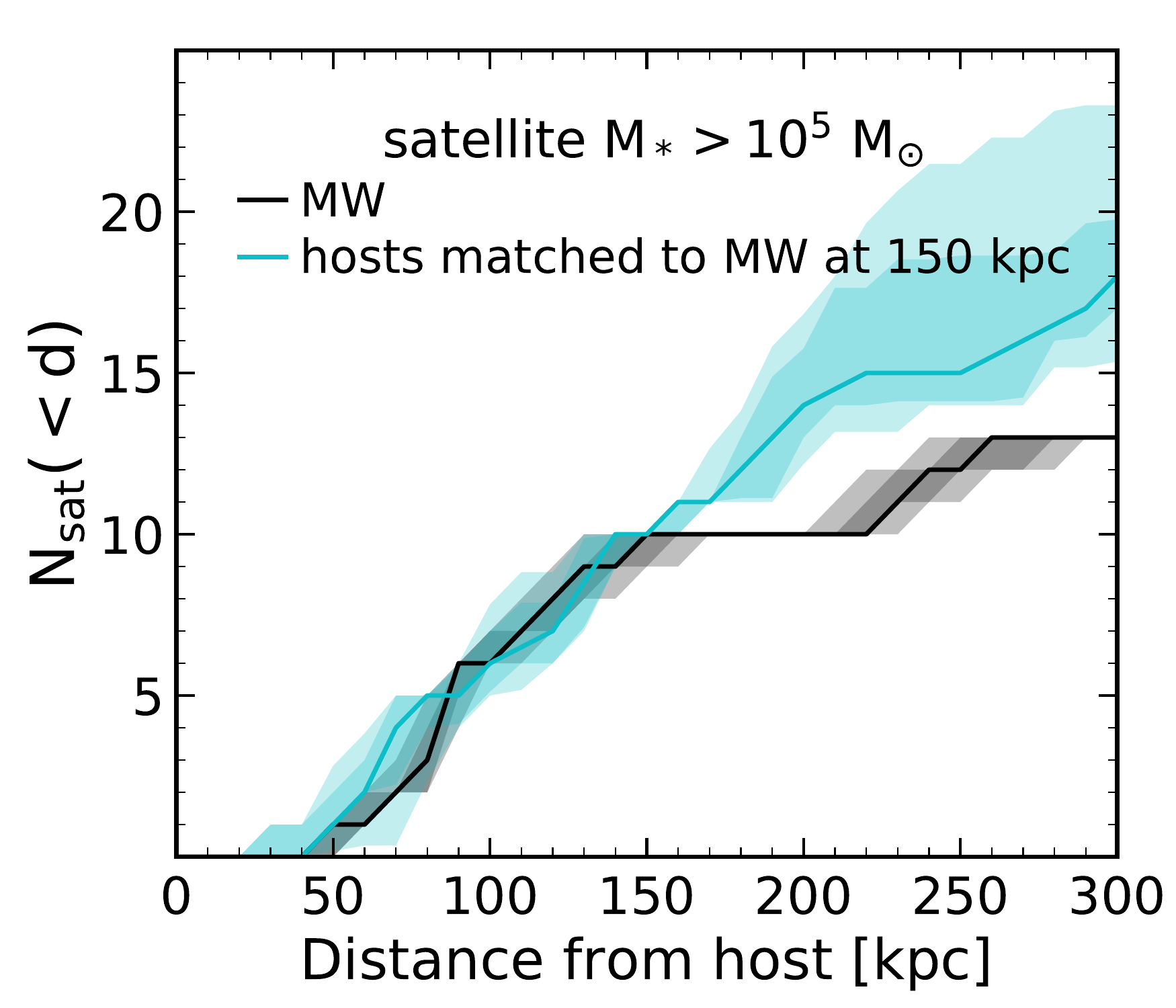}\par
	\includegraphics[width=\columnwidth]{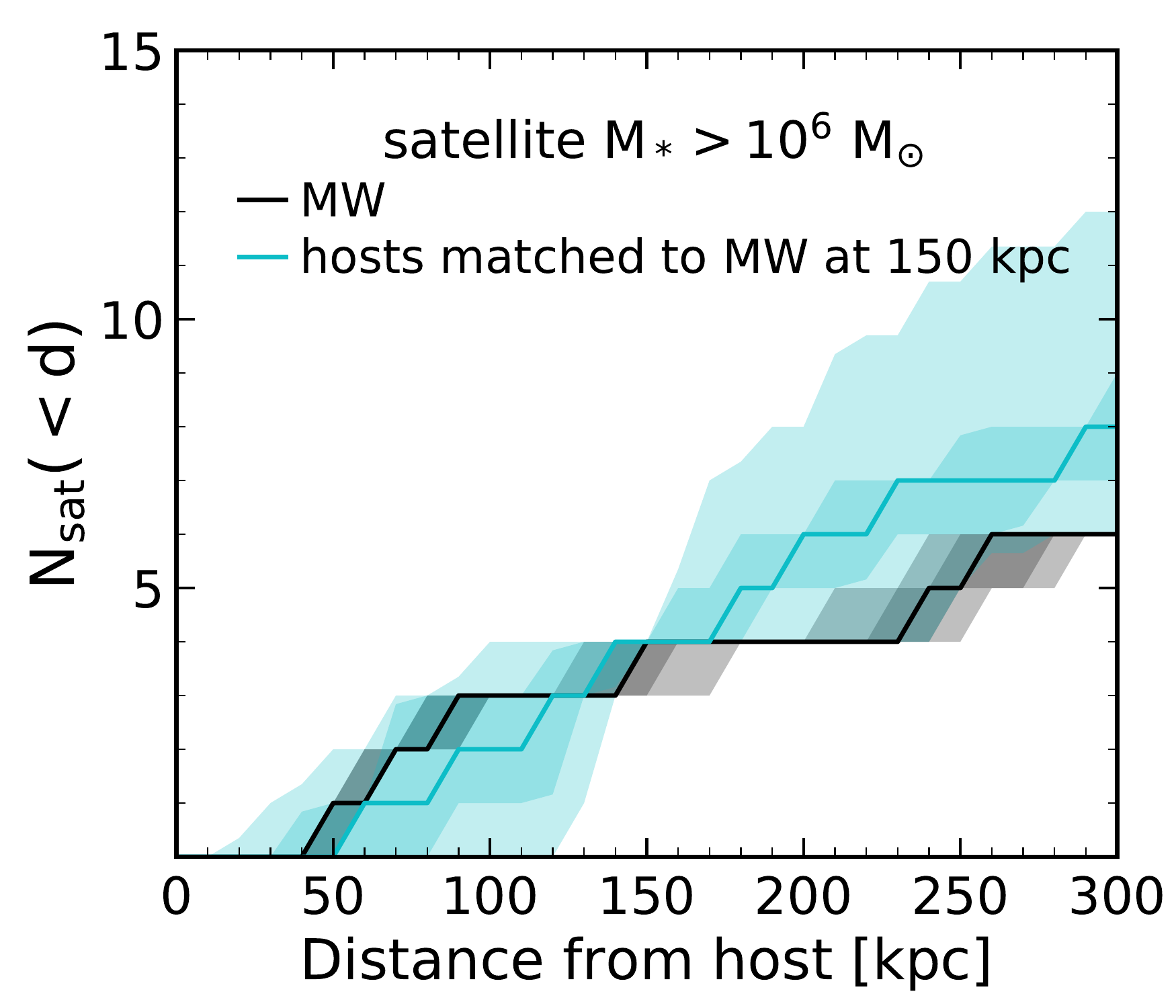}\par
	\end{multicols}
	\vspace{-7 mm}
    \caption{Comparison of the MW with simulated profiles matched to it at 150 kpc, and implications for incompleteness of satellite galaxies around the MW. Note that we compare our simulations to observations of the MW that have \textit{not} been completeness-corrected.
    \textit{Left:} The cumulative number of satellite galaxies with M$_* > 10^5$ M$_{\odot}$ as a function of 3D distance from the host for the MW (black) and simulations (blue) that match the number of satellites around the MW within 150 kpc. We find 8 profiles across all hosts and snapshots that meet this criteria. Within 150 kpc the agreement between the simulations and observations is excellent, but beyond 150 kpc all the simulations lie systematically at least 2 (and more commonly 5) satellites above the observations. This indicates that observations of the MW may be incomplete for satellites with M$_* > 10^5$ M$_{\odot}$.
    \textit{Right:} Same as left, but for satellite galaxies with M$_* > 10^6$ M$_{\odot}$. We find 27 profiles across all hosts and snapshots that match the MW at this mass limit. The agreement between the simulations and observations spans the full distance range for this mass bin. Though the MW lies within simulation scatter, the simulation median is 1-2 satellites higher than the MW beyond 150 kpc. Observations of the MW are likely complete or nearly complete for satellite galaxies with M$_* > 10^6$ M$_{\odot}$ based on our simulations.
    }
    \label{mw_prediction}
\end{figure*}
%%% MW PREDICTIONS FIGURE %%%

\subsection{Implications for incompleteness around the Milky Way}\label{MW_inc}

While the MW and M31 profiles agree quite well out to 150 kpc, the MW appears to have a larger proportion of its satellite galaxies at small distances than both our simulations and M31.
This could be a peculiarity of the MW profile, or it may be hinting at more satellites remaining to be discovered beyond 150 kpc from the MW.
For example, the difference in shape could be due to the current presence of the LMC and the SMC near their pericenters around the MW \citep{Kallivayalil2013}.
While \citet{Yniguez2014} found that potential incompleteness in the census of MW satellites meant there could be $\sim$10 classical dwarf satellite galaxies remaining to be discovered, which could bring the MW into better agreement with M31.

We expect observations of MW satellites to be complete down to at least M$_* \sim 10^5$ M$_{\odot}$ within 150 kpc and out of the plane of the disk.
Beyond this distance and through the disk the completeness may be uncertain, as evidenced by the discovery of Antlia 2, which had been obscured by the MW disk.
Here, we focus on implications for incompleteness without considering the effects of seeing through the MW's disk.
While our theoretical results are suggestive, a more in-depth account of observational completeness for `classical' dwarf galaxies also depends on the surface brightness distribution of the population and their on-the-sky positions with respect to the Galactic plane (or any other foreground structure). 
Our simulations can provide more detailed predictions for these effects on the completeness of the satellite population, especially through the use of Gaia-like mocks \citep{Sanderson2018}, which we plan to pursue in future work.

To investigate potential incompleteness in observations of the MW's satellites (that have \textit{not} been completeness-corrected), we examine how many additional satellites we would expect to find around the MW based on our simulations that match the MW profile out to 150 kpc.
We choose simulated profiles for comparison by requiring them to have the same number of satellites within 150 kpc as the median value for the MW, which is 10 for M$_* > 10^5$ M$_{\odot}$.
One host meets this criteria at four snapshots (m12z), and four hosts meet this criteria at a single snapshot each (m12w, m12r, Romeo, and Juliet), providing a total of 8 matched profiles.

Figure~\ref{mw_prediction} (left) shows the range of simulated profiles that match the MW at 150 kpc compared to the observed MW profile, for satellites with M$_* > 10^5$ M$_{\odot}$.
The simulations agree remarkably well with the MW below 150 kpc, which further strengthens our claim that if we match the profile at this distance, then we are accurately resolving survivability of satellites closer to the host.
Notably, beyond $\sim$150 kpc the simulation profiles are systematically higher than the MW profile.
In total, the simulation median profile has 5 more satellites than the MW median profile within 300 kpc.
The lower 68 per cent (95 per cent) limits on the simulation profile imply that there may be at least 4 (2) more satellites at 150-300 kpc from the MW. 
If our simulations are representative of the real MW, then based on the median simulation profile, we predict that there should be 5 more satellites with M$_* > 10^5$ M$_{\odot}$ within 150-300 kpc of the MW.

We expect observational completeness to be better at higher satellite stellar masses, so we repeat this exercise for satellites with M$_* > 10^6$ M$_{\odot}$ to check if the agreement between simulations and observations is indeed better.
At this satellite stellar mass threshold, the MW has 4 satellites within 150 kpc.
We find that 8 out of the 12 simulated hosts match the MW's profile at 150 kpc for at least 1 snapshot out of 11, together providing a total of 27 matching profiles.
Notably, m12i matches for 8 snapshots and Romeo matches for 5 snapshots.

Figure~\ref{mw_prediction} (right) shows the range of simulated profiles that match the MW at 150 kpc compared to the observed MW profile, for satellites with M$_* > 10^6$ M$_{\odot}$.
We find that the agreement between the simulations and observations spans the full range of distances in this satellite mass range.
The 95 per cent simulation scatter almost completely encompasses the MW observational scatter below 150 kpc, and beyond that the 95 per cent simulation scatter overlaps with the upper half of the observational scatter.
The lower 68 per cent (95 per cent) limits on the simulation profile imply that there may be at least 1 (0) more satellite with M$_* > 10^6$ M$_{\odot}$ to be discovered within 150-300 kpc of the MW.
The median simulation profile indicates that there are on average 2 satellites in this mass range remaining to be discovered around the MW.
Compared to the larger number of undiscovered satellites that we predict for the lower mass range, the observations of satellites with M$_* > 10^6$ M$_{\odot}$ appear to be more complete. 
We find that this strengthens our conclusion that the census of MW satellite galaxies may not be complete down to M$_* > 10^5$ M$_{\odot}$.

%%% MOCK PANDAS FIGURE %%%
\begin{figure*}
    \begin{multicols}{2}
	\includegraphics[width=\columnwidth]{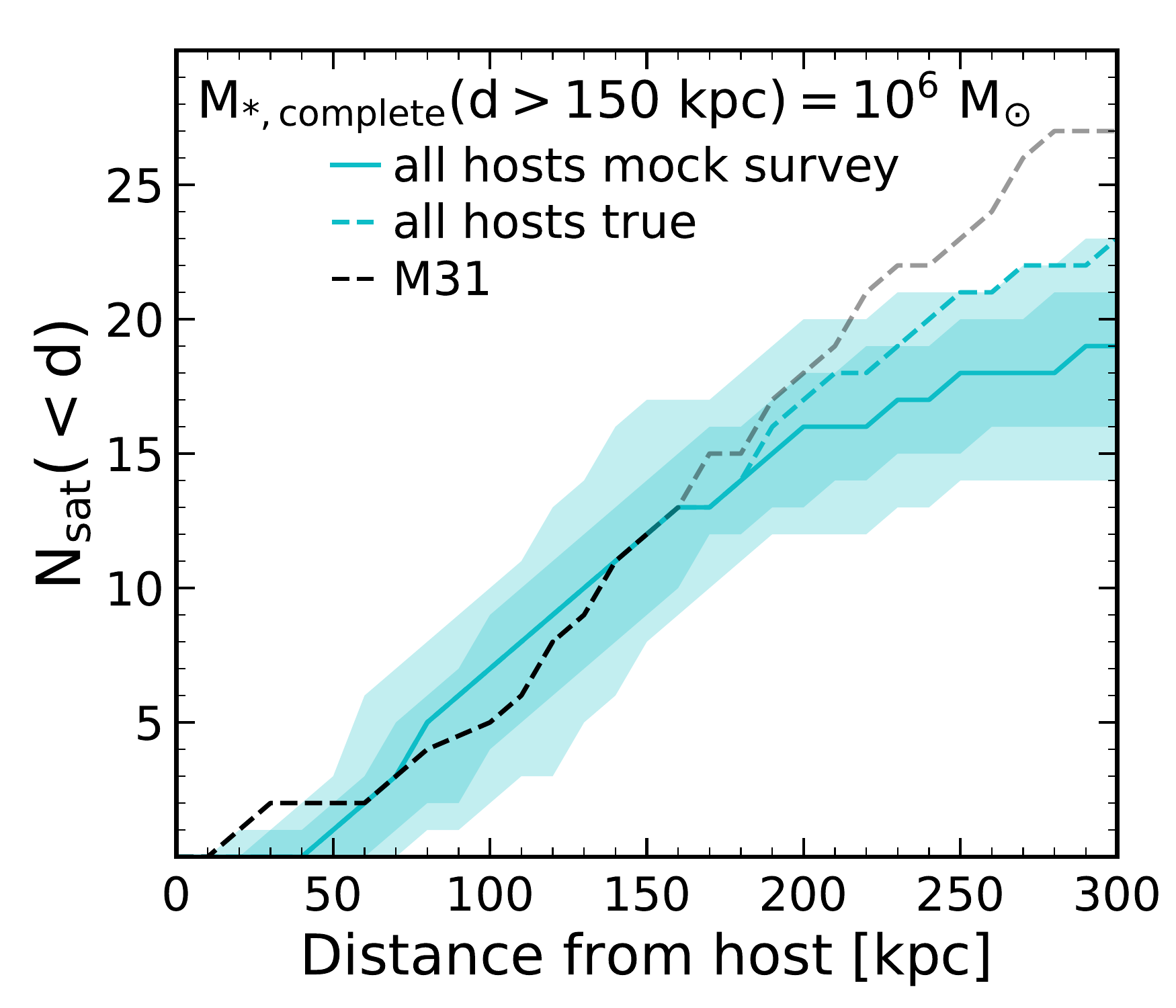}\par
	\includegraphics[width=\columnwidth]{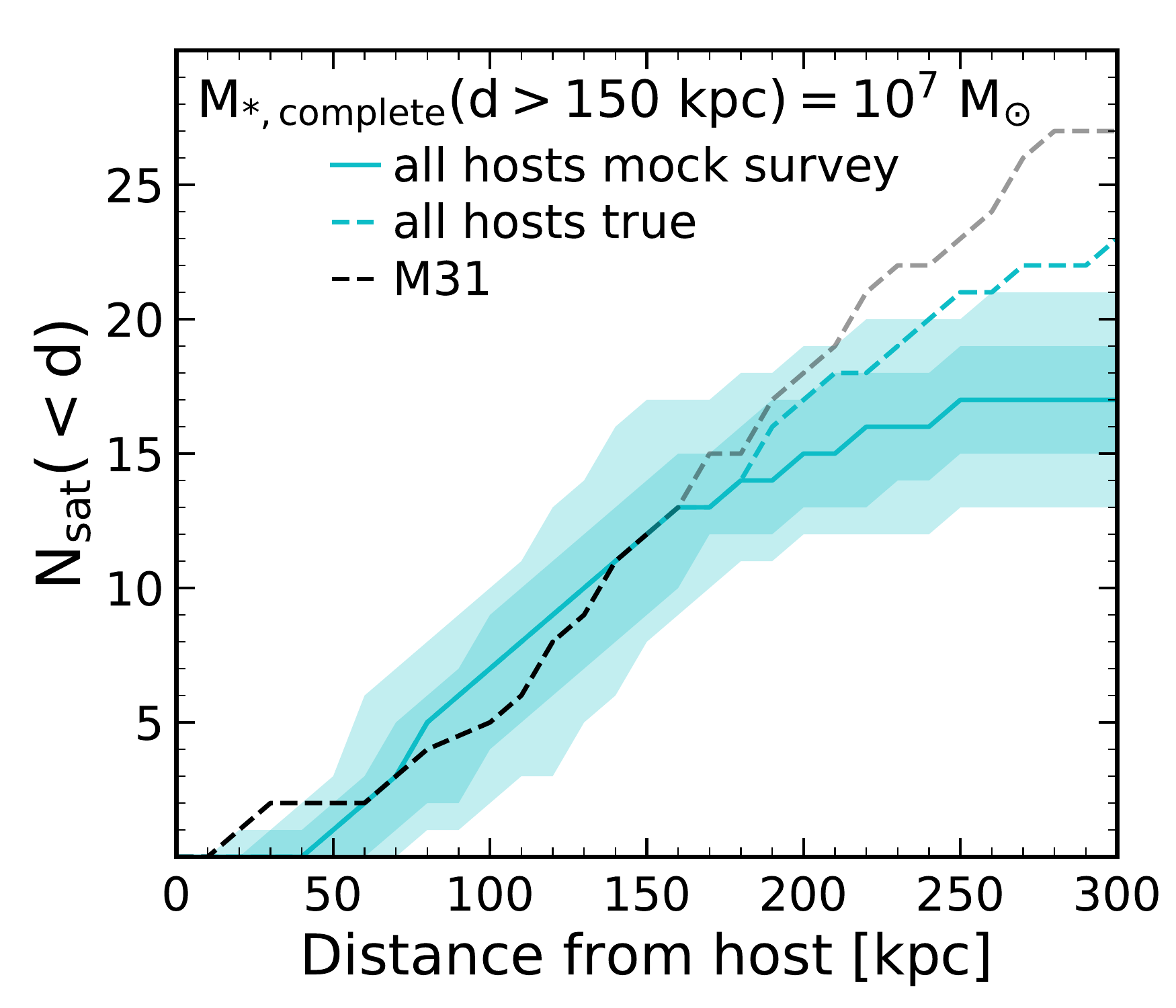}\par
	\end{multicols}
	\vspace{-7 mm}
    \caption{Implications for incompleteness of M31 satellites as a function of distance. M31's line is lighter where the observational data are known to be incomplete.
    \textit{Left:} Solid blue line and scatter shows the radial profile measured by a mock survey that is complete to M$_* = 10^5$ M$_{\odot}$ within 150 kpc (to mock the PAndAS footprint) and complete to M$_* = 10^6$ M$_{\odot}$ for 150-300 kpc. Dashed blue line shows the true radial profile for all satellites with M$_* > 10^5$ M$_{\odot}$. Dashed black line shows M31's profile for comparison. Incompleteness causes the mock survey to miss $\sim$20 per cent of the satellites.
    \textit{Right:} Same as left, but for a mock survey that is complete to M$_* = 10^7$ M$_{\odot}$ within 150-300 kpc. Here, incompleteness causes the mock survey to miss $\sim$25 per cent of the satellites. If our simulations are accurate representations of M31-like satellite populations, these results predict that there are 6-9 satellite galaxies to discover around M31.
    }
    \label{mock_pandas}
\end{figure*}
%%% MOCK PANDAS FIGURE %%%

\subsection{Incompleteness around M31}\label{M31_inc}

M31's satellite population is complete down to our lowest stellar mass limit (M$_* > 10^5$ M$_{\odot}$) and within 150 kpc of the host given the uniform depth and coverage of PAndAS in this area (\citealt{McConnachie2009} and see Section~\ref{observations} for more discussion). % EJT: I made a minor text change here of just moving where the citation is
However, outside of the PAndAS footprint, the completeness limit for M31's satellite galaxies is not clear.
We use our simulations as testing grounds to examine effects of this incompleteness on recovering M31's true radial profile.
For simplicity and to match M31's profile (which has a median value of 27 satellites at 300 kpc), we select hosts from our simulations that have at least 20 (median over time) satellite galaxies within 300 kpc with M$_* > 10^5$ M$_{\odot}$: m12m, m12c, m12w, Juliet, and Louise. 
We perform a mock survey by selecting satellites in 2D projection along 1000 lines of sight. 
To mimic the PAndAS footprint, we assume that our mock observations are complete down to M$_* = 10^5$ M$_{\odot}$ within a projected radial distance of 150 kpc from the host, and within 150-300 kpc we assume two possible estimates of the completeness: M$_* > 10^6$ M$_{\odot}$ and M$_* > 10^7$ M$_{\odot}$.

Figure~\ref{mock_pandas} shows the results of our mock surveys compared to the true radial profiles for the 5 hosts with M31-like profiles.
Comparing our simulated true median profiles (blue dashed) to the recovered profiles (blue solid, with shaded regions showing 68 per cent and 95 per cent scatter), we find that we typically recover 75-80 per cent (median) of our satellites, depending on the completeness mass. Thus, if our estimates of stellar completeness beyond 150 kpc are correct, M31 reasonably has 6-9 undetected satellites with M$_* > 10^5$ M$_{\odot}$ within 150-300 kpc of the host, which we obtain by applying 20-25 per cent incompleteness to M31's observed profile.
It is also worth noting that beyond $\sim$200 kpc, the M31 profile lies above the scatter in the selected simulations.
This may indicate that M31 is more massive than our simulated hosts, or that there is something else fundamentally different about M31 compared to our simulations.
This result motivates deeper PAndAS-like surveys out to greater distances around M31, which are likely to find several dwarf galaxies, based on our simulations.

\section{Summary and Discussion}\label{conclusion}

Using the FIRE-2 baryonic cosmological zoom-in simulations of MW- and M31-mass halos, we study the radial profiles of satellite galaxies with M$_* > 10^5$ M$_{\odot}$. 
We explore 12 host-satellite systems: 8 isolated MW/M31-like galaxies from the Latte suite + m12z and 4 galaxies in LG-like pairs from the ELVIS on FIRE suite, where the hosts span M$_{200\rm{m}}=0.9-1.7\times10^{12}$ M$_{\odot}$.
To reduce noise in profiles at small distances from satellites momentarily near pericenter, we time-average the simulated radial profiles over $z=0-0.1$ ($\sim$1.3 Gyr).
We compare against the 3D profiles measured around the MW and M31 (including observational uncertainties in line-of-sight distance), and against the 2D profiles of MW analogs in the SAGA survey.
Our main conclusions are as follows:

\begin{itemize}
    \item The radial distributions of satellite galaxies with M$_* > 10^5$ M$_{\odot}$ within 300 kpc of their host in the FIRE-2 simulations agree well with LG observations. The scatter in the simulations spans the radial profiles of the MW and M31, and the median ratio of simulated-to-observed profiles is typically $\sim$1 for the MW and $\sim$1/2 for M31. Though M31 has a relatively large satellite population, it is still within our simulation scatter.
    \item The radial concentration of the baryonic simulations generally agrees with LG observations, but the MW (and M31 in 2D projection) has a more concentrated shape than the simulations. If we examine simulations with the same number of satellites as the MW at d$<$150 kpc, we find excellent agreement with the MW down to $\sim$50 kpc. Beyond 150 kpc, the matched simulation profiles all lie above the MW profile. We predict 2-10 satellites (at 95\%) with M$_* > 10^5$ M$_{\odot}$ to be discovered within 150-300 kpc from the MW.
    \item If we perform mock surveys with the same observational characteristics as PAndAS on our simulations, we recover on average 75-80 per cent of the true satellite population. Based on this, we predict there may be 6-9 undetected satellites around M31 and outside the PAndAS footprint depending on the (uncertain) completeness limit outside of the PAndAS footprint.
    \item 2D projected radial profiles of satellite galaxies with M$_* > 5 \times 10^6$ M$_{\odot}$ for the simulations also agree with the profiles for the 8 MW analogs from the SAGA survey. The scatter in the simulations spans a majority of SAGA profiles, though 3 SAGA hosts have fewer satellites at large distances ($>$100 kpc).
    \item The agreement we find in radial profiles does not depend strongly on \textit{satellite} galaxy stellar mass. Thus, even at small distances (<100 kpc) where satellite galaxies are subject to stronger tidal forces from the host's disk, our simulations resolve the survival and physical destruction of satellites down to our lower stellar mass limit (M$_* > 10^5$ M$_{\odot}$, with typical M$_{\rm peak} > 8 \times 10^8$ M$_{\odot}$ or $\sim 2 \times 10^4$ DM particles).
    \item Simulated \textit{hosts} with larger stellar masses have fewer satellite galaxies at small distances ($\lesssim100$ kpc). We interpret this as caused by tidal destruction of satellite galaxies by the gravitational potential the host's disk. We find a similar correlation with halo mass as well, which we interpret as a manifestation of more massive halos having bigger disks. We note, however, that we examined hosts only over a narrow host halo mass range M$_{\rm h} = 0.9 - 1.7 \times 10^{12}$ M$_{\odot}$.
    \item The variation from host-to-host scatter among the different simulations dominates over time variation at large distances ($\gtrsim$100 kpc), while time variation is the dominant contributor to scatter at small distances ($\lesssim$50 kpc).
    \item KS tests between the radial profiles of the simulations the profiles of the MW and M31 show that most of the simulated profiles are consistent with being drawn from the same underlying distribution as LG observations. However, 4 (1) of the simulations have radial profiles inconsistent with the MW (M31).
    \item Consistent with previous studies, our dark matter-only simulations have many more subhalos at small distances (<100 kpc), and hence larger concentrations in their radial profiles, than their baryonic counterparts. This corroborates the idea that the baryonic simulations have enhanced tidal destruction of satellites due to the additional disk potential present in baryonic hosts. We provide fits to the ratio of baryonic to dark matter-only subhalo counts as a function of distance, which one can use to renormalize existing DMO simulations to include baryonic effects.
\end{itemize}

We present a thorough comparison of satellite galaxy radial profiles around MW/M31-like galaxies in the FIRE-2 simulations to the LG and to MW analogs from the SAGA survey.
Incorporating time dependence of the radial profile over the last 1.3 Gyr in the simulations is key to a robust comparison of the simulations with observations, because the profile at small distances from the host can be highly time-variable.
Overall, we find that our simulations are generally representative of current observations.
Specifically, combined with the recent results of \citet{GK2018} and \citet{GK2019}, who analyzed the same FIRE simulation suite, we see broad agreement with the population of `classical' dwarf galaxies (M$_* \gtrsim 10^5$ M$_\odot$) in the LG across a wide range of properties: stellar masses, stellar velocity dispersion and dynamical mass profiles, star-formation histories, and now spatial distributions in terms of radial profiles.
However, we emphasize that these simulations are not yet able to resolve ultra-faint dwarf galaxies, so it remains unclear how well simulations agree with the profiles of ultra-faints (especially in incorporating incompleteness).
The simulations used here also do not yet include the most realistic treatment of cosmic ray physics implemented in the FIRE project, which has effects on the mass of the host and hence the survivability of satellites \citep[][Hopkins et al., in preparation]{Chan2018}.

%Sampling simulated hosts over a range in halo mass that spans the uncertainties of the LG halo masses also allows us to examine the radial profiles as a function of host properties as well.
%We find evidence for baryonic effects on the radial profile of satellites at small distances, where host effects are strongest. 
%In particular we find that hosts with more massive stellar disks have fewer satellites within both 50 and 100 kpc of the host. 
%We find similar trends with host halo mass within 100 kpc, but within 50 kpc halo mass is no longer strongly correlated with the number of satellites. 
%To further explore the influence of the baryonic disk, we scale the baryonic radial profiles to the dark matter-only profiles. 
%The median ratio of baryonic-to-DMO subhalos is flat at large distances from the hosts (>200 kpc) and rapidly declines to zero within 10-15 kpc from the hosts. 

Given the correlation with number of satellites at small distances with host stellar mass, we interpret the analogous dearth of subhalos in the baryonic simulations relative to the DMO simulations as primarily from tidal disruption of satellites by the baryonic disk.
This agrees with a wealth of previous work that generally finds an excess of DMO subalos near the host relative to the number in baryonic simulations or DMO+analytical disk potential \citep{Taylor2001,Hayashi2003,Read2006a,Read2006b,Berezinsky2006,DOnghia2010,Penarrubia2010,Brooks2013,Zhu2016,Errani2017,GK2017,Sawala2017,Kelley2018}.

The shape of the radial profile of satellite galaxies also has significant implications for how other satellite phenomena are measured.
For example, the missing satellites problem \citep[e.g.][]{Moore1999,Klypin1999} and the satellite plane problem \citep[e.g.][]{Pawlowski2018} are both sensitive to concentration of the radial profile, and the MW's satellite distribution is often found to be unusually concentrated compared to simulations \citep[e.g.][]{Zentner2005,Li2008,Metz2009,Yniguez2014}.
However, controlling for the shape of the profile proves difficult because typical metrics of radial concentration do not necessarily produce the comprehensive description of spatial distribution that is needed to interpret observations.
We find that DMO simulations have systematically higher radial concentration than baryonic simulations.
Other studies have reached the same conclusion by comparing DMO simulations to baryonic simulations or to DMO simulations with a semianalytic model of galaxy formation \citep[e.g.][]{Kang2005,Ahmed2017}.
This suggests that DMO simulations alone cannot be used to accurately predict the shapes of observed radial profiles which are likely affected by baryonic processes.

We also find that while the radial concentration of the M31 profile agrees with our baryonic simulations, the MW is more concentrated than the baryonic simulations when we compare profile shape with $R_{90}/R_{50}$. 
The MW is more concentrated than the baryonic simulations (and even most of the DMO simulations) under this metric because 50 per cent of MW satellites are within 110 kpc of the MW, but the simulations only attain this fraction of satellites within $\gtrsim$140 kpc of the host on average. 
This is similar to what \citet{Yniguez2014} found by comparing the number of satellites within 100 and 400 kpc of their host for LG profiles and DMO simulations: the MW has a more concentrated shape than all of their simulations and M31. 
If we instead match the number of simulated satellites within 150 kpc of their host to the observed number within 150 kpc of the MW, we find that simulations meeting this criteria unanimously show a larger number of satellites within 150-300 kpc than the MW.
We interpret this as potential evidence for incompleteness in the MW's satellite population at large distances, and our simulations predict there are on average 5 (at least 2) satellites with M$_* > 10^5$ M$_{\odot}$ to be discovered beyond 150 kpc from the MW.

Due to the peculiarity of the MW profile, we also use KS testing to accurately compare our simulations with observations. 
We find that all 12 of the simulated hosts have at least 10 snapshots matching M31's profile, and 9 of the hosts have at least 10 snapshots matching the MW's profile.
We will examine the full three-dimensional spatial and dynamical distributions of satellite galaxies in detail and examine the satellite plane problem in our simulations in future work (Samuel et al., in preparation).

The spatial distribution of satellite galaxies correlates with attributes of the host galaxy, both in our simulations and in the LG.
Importantly, the correlated host attributes are not limited to the dark matter halo properties of the host, and the spatial distribution of satellites may be most strongly correlated with the host's baryonic features.
DMO simulations are insensitive to the effects of a realistic host galaxy disk, and thus are not sufficient predictors of observed radial profiles at small distances which are the most influenced by the host's baryonic structure.
We have provided a correction to such DMO radial profiles by modeling the depletion of subhalos by the baryonic disk as a function of distance from the host.

\section*{Acknowledgements}

We thank Marla Geha, Risa Wechsler, and Ethan Nadler for their helpful comments. This research made use of Astropy,\footnote{http://www.astropy.org} a community-developed core Python package for Astronomy \citep{astropy:2013, astropy:2018}, the IPython package \citep{ipython}, NumPy \citep{numpy}, SciPy \citep{scipy}, Numba \citep{numba}, and matplotlib, a Python library for publication quality graphics \citep{matplotlib}.

JS, AW, and SB were supported by NASA, through ATP grant 80NSSC18K1097 and HST grants GO-14734 and AR-15057 from the Space Telescope Science Institute (STScI), which is operated by the Association of Universities for Research in Astronomy, Inc., for NASA, under contract NAS5-26555.
We performed this work in part at the Aspen Center for Physics, supported by NSF grant PHY-1607611, and at the KITP, supported NSF grant PHY-1748958.
Support for SGK and PFH was provided by an Alfred P. Sloan Research Fellowship, NSF grant \#1715847 and CAREER grant \#1455342, and NASA grants NNX15AT06G, JPL 1589742, 17-ATP17-0214.
Support for SRL was provided by NASA through Hubble Fellowship grant \#HST-JF2-51395.001-A awarded by STScI.
KE was supported by an NSF graduate research fellowship.
MBK acknowledges support from NSF grant AST-1517226 and CAREER grant AST-1752913 and from NASA grants NNX17AG29G and HST-AR-13888, HST-AR-13896, HST-AR-14282, HST-AR-14554, HST-AR-15006, HST-GO-12914, and HST-GO-14191 from STScI.
CAFG was supported by NSF through grants AST-1517491, AST-1715216, and CAREER award AST-1652522, by NASA through grant 17-ATP17-0067, and by a Cottrell Scholar Award from the Research Corporation for Science Advancement.
JSB was supported by NSF AST-1518291, HST-AR-14282, and HST-AR-13888.
We ran simulations using the Extreme Science and Engineering Discovery Environment (XSEDE) supported by NSF grant ACI-1548562, Blue Waters via allocation PRAC NSF.1713353 supported by the NSF, and NASA HEC Program through the NAS Division at Ames Research Center.

%%%%%%%%%%%%%%%%%%%%%%%%%%%%%%%%%%%%%%%%%%%%%%%%%%


\bibliographystyle{mnras}
\bibliography{radial}

%%%%%%%%%%%%%%%%%%%%%%%%%%%%%%%%%%%%%%%%%%%%%%%%%%

%%%%%%%%%%%%%%%%% APPENDICES %%%%%%%%%%%%%%%%%%%%%

\appendix

\section{Resolution test}\label{lowres_appendix}

To examine the dependence of our satellite profiles on numerical resolution, we use the lower-resolution (LR) versions of the Latte simulation suite. 
We simulated each of the 7 Latte hosts at 8$\times$ lower mass resolution, with baryonic particle masses of m$_{\rm bary}\sim5.7\times10^4$ M$_{\odot}$ and m$_{\rm dm}=2.8\times10^5$ M$_{\odot}$. 
Furthermore, all gravitational force softenings are 2$\times$ larger.

%Given the larger particle masses, selecting LR satellites based on the same number of star particles (20) in a high resolution (HR) satellite of M$_{*}\sim10^5$ M$_{\odot}$ corresponds to a stellar mass of M$_{*}\sim8\times10^5$ M$_{\odot}$ in the LR simulations.
In principle, we could compare satellites at fixed M$_*$ between LR and high-resolution (HR) simulations.
However, we choose to compare the survival of \textit{subhalos} in the baryonic simulations that are resolved with the same number of DM particles, for three reasons.
First, as studied extensively in Section 4.1.4 of \citet{Hopkins2018}, the stellar masses of dwarf galaxies resolved with small numbers of star particles are sensitive to numerical convergence; the lowest-mass galaxies resolved in our LR simulations ($\sim 20$ star particles) form systematically $\sim2\times$ higher M$_*$ at fixed subhalo M$_{\rm peak}$ than in our HR simulations.
Thus, comparing satellites at fixed M$_*$ at our resolution limit mixes the numerical effects of star-formation efficiency and tidal disruption, but comparing satellites at fixed M$_{\rm peak}$ isolates the effects of tidal disruption, which is our goal here.
Second, because dwarf galaxies at these masses are so DM-dominated, the survivability of a satellite is governed more directly by the number of DM particles in its subhalo than its number of star particles.
Finally, most previous works on numerical disruption of satellites \citep[e.g.][]{vandenBosch2018} focused on DM-only simulations and how well resolved subhalos are, so using M$_{\rm peak}$ makes our tests more comparable to those previous works.

%In principle we would then compare LR and HR satellites at the same stellar mass limit, but we find that dwarf galaxies in our LR simulation form systematically $\sim2\times$ more stellar mass (at fixed halo mass), as \citet{Hopkins2018} also discussed.
%Thus, making a direct comparison of LR to high resolution (HR) at fixed stellar mass difficult to interpret due to baryonic effects and we instead choose to use (sub)halo peak mass limits that correspond to these stellar mass limits.

%Thus, in lieu of comparing satellite galaxies at fixed M$_*$, we compare \textit{subhalos} that are resolved with the same number of DM particles as a galaxy with M$_{*} \sim 10^5$ M$_{\odot}$ in the HR simulations.
%We found subhalos hosting galaxies with M$_{*} \sim 10^5$ M$_{\odot}$ in the HR simulations and measured their peak masses over time to be
%We found subhalos hosting galaxies with M$_{*} \sim 10^5$ M$_{\odot}$ in the HR simulations and measured their peak masses over time to be M$_{\rm{peak}}\sim8\times10^8$ M$_{\odot}$.
For the lowest-mass galaxies that we examine in this work, M$_{*} \sim 10^5$ M$_{\odot}$, we find that, across the HR simulations, they are hosted by subhalos with average M$_{\rm{peak}}\sim8\times10^8$ M$_{\odot}$ ($\sim2\times10^4$ DM particles).
Similarly, the HR satellites with M$_{*} \sim 10^6$ M$_{\odot}$ have subhalos of M$_{\rm{peak}}\sim2\times10^9$ M$_{\odot}$ ($\sim7\times10^4$ DM particles) and HR satellites with M$_{*} \sim 10^7$ M$_{\odot}$ have subhalos of M$_{\rm{peak}}\sim10^{10}$ M$_{\odot}$ ($\sim3\times10^5$ DM particles).
With 8$\times$ larger particle mass in the LR simulations, a LR subhalo that is resolved as well as the subhalos of our lowest-mass satellites in the HR simulations has M$_{\rm peak}\sim6.4\times10^9$ M$_{\odot}$.
We thus select subhalos in both the LR and HR simulations with  M$_{\rm{peak}}>6.4\times10^9$ M$_{\odot}$ and compute their radial profiles out to 1000 kpc around each host, averaging over all 67 snapshots at $z=0-0.1$ to improve statistics.

\begin{figure}
	\includegraphics[width=\columnwidth]{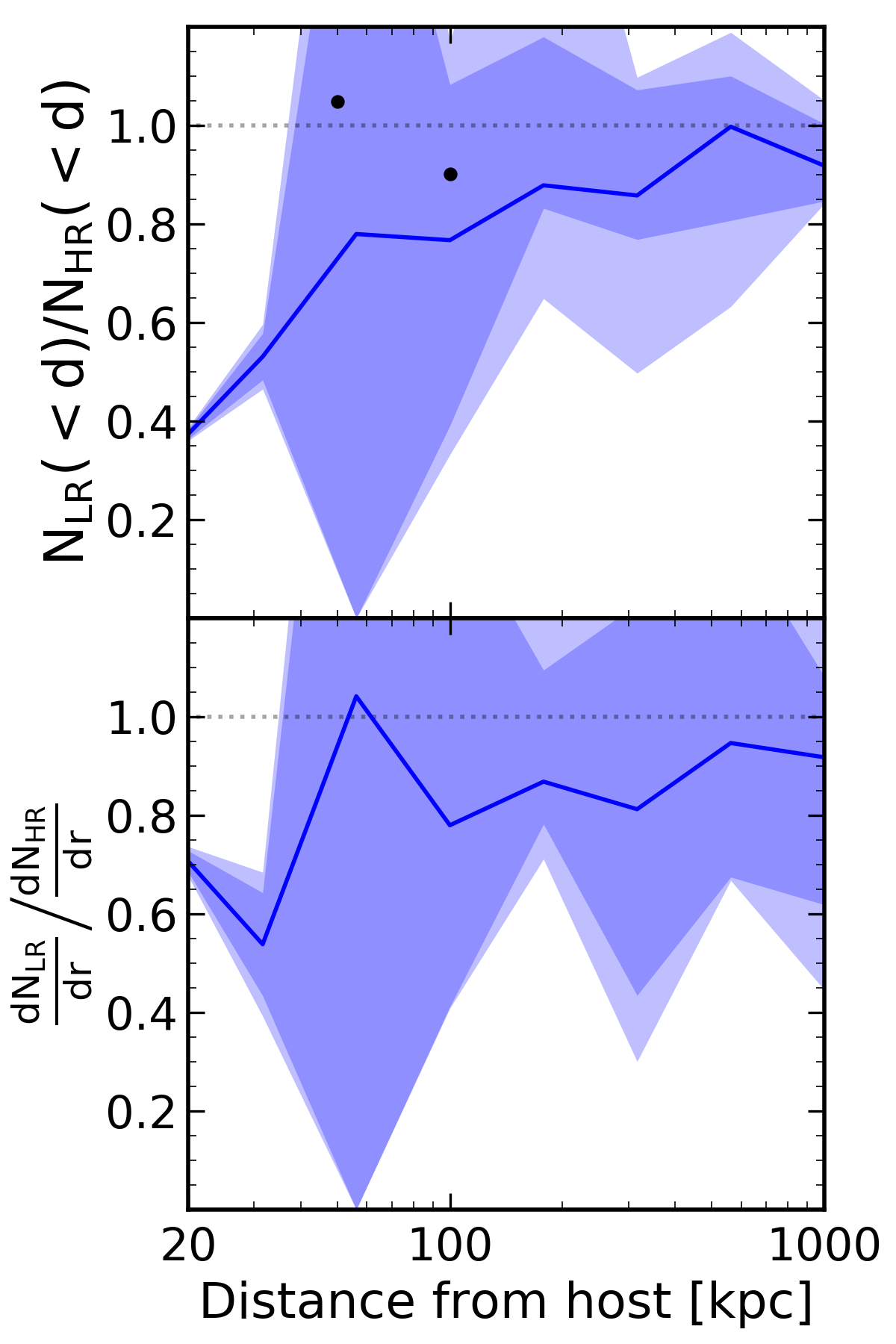}
	\vspace{-7 mm}
    \caption{Resolution test for subhalos with M$_{\rm peak}>6.4\times10^9$ M$_{\odot}$, which have the same number of DM particles in the low-resolution (LR) simulation as the lowest-mass subhalos that we analyze in the high-resolution (HR) simulations. \textit{Top:} The cumulative radial profiles for subhalos in the LR simulations normalized to the HR profiles. The blue line shows the median and the shaded regions show the host-to-host scatter. The LR simulations have on average 80-100 per cent the number of subhalos as the HR simulations beyond about 30 kpc, indicating that we are resolving the satellites in our lowest mass bin from the main text. At small distances, the LR simulations have significantly fewer subhalos than the HR versions, but this is caused at least in part because of the more massive baryonic disks of the LR hosts. The black points represent an approximate model for removing this host mass effect at 50 and 100 kpc using the fits from Figure~\ref{nsat_v_host_mass}.
    \textit{Bottom:} Same as top, but for differential radial binning instead of cumulative.}
    \label{lowres}
\end{figure}

Figure~\ref{lowres} shows a comparison of the LR radial profiles to the HR versions.
The top panel compares the cumulative profile while the bottom panel compares the differential (discrete distance bins) profile, which more directly indicates where converge occurs.
The solid lines and shaded regions show the median and 68 per cent and 95 per cent host-to-host scatter over the 7 Latte simulations.
Within $\sim$30 kpc, the LR simulations show a deficit of $\sim$25 per cent (in the differential) compared with HR.
Beyond 30 kpc, the median ratio stays mostly between 0.8 and 1, and the 68 per cent host-to-host scatter is always consistent with 1 for both cumulative and differential profiles.
We thus conclude that the radial profiles are well converged (to $\sim20$ per cent) beyond $\sim$30 kpc, where almost all observed satellites (M$_{*}>10^5$ M$_{\odot}$) of the MW and M31 are.
%This test shows that at distances $\gtrsim30$ kpc, LR subhalos resolved with the same number of DM particles as HR subhalos hosting M$_{*}\sim10^5$ M$_{\odot}$ galaxies are converged with respect to the HR radial profile.

Moreover, the difference between LR and HR within 30 kpc is exaggerated by the fact that the LR simulations have more massive host galaxies.
As with the dwarf galaxies themselves, and as studied in detail in \citet{Hopkins2018}, the stellar masses of the host galaxies are sensitive to resolution as well, with LR hosts having on average 1.7 times higher stellar mass than their HR counterparts.
Using the results from Figure~\ref{nsat_v_host_mass} (left), a host galaxy with 1.7$\times$ higher stellar mass will have $\sim$15 per cent fewer satellites at d$<$50 kpc, even at fixed resolution, which can account for most of the difference between LR and HR at small distances.
Based on this, we plot the expected median values for the ratio of LR to HR profiles at 50 and 100 kpc as black points in the upper panel of Fig~\ref{lowres}.
These profile ratio values are within 10 per cent of unity, showing better agreement between the LR and HR simulations.

Furthermore, given the complicating effects of different host galaxy masses in LR versus HR simulations, we also note similar results from the extensive numerical tests in \citet{Hopkins2018}.
Specifically, in Section 4.14 and Figure 14 they compared the (differential) number of subhalos versus distance in a DM-only simulation of the same m12i host at the same resolutions that we use here, using a broadly similar subhalo selection (instantaneous M$_{\rm bound} > 10^8$ M$_\odot$).
Thus, while that convergence test did not include the additional tidal force of the central galaxy or any other baryonic effects, its does provide a cleaner numerical test in the DM-only regime.
They found convergence to better than 20 per cent down to $d \approx 50$ kpc, consistent with the results of Figure~\ref{lowres}.

We also can use the more rigorous criteria articulated in \citet{vandenBosch2018} to test how well our lowest-mass subhalos are resolved.
They stipulate that a subhalo on a circular orbit around a static, spherically symmetric host potential will suffer from numerical noise or disruption if the bound mass fraction falls below either of two limits that depend on mass and force resolution:

\begin{equation}
f_{\rm bound}<0.32(N_{\rm acc}/1000)^{-0.8}
\end{equation}

\begin{equation}
f_{\rm bound}<\frac{1.79}{f(c)}\left(\frac{\epsilon}{r_{\rm s,0}}\right)\left(\frac{r_{\rm h}}{r_{\rm s,0}}\right)
\end{equation}

Where $N_{\rm acc}$ is the number of DM particles in the subhalo at accretion, $c$ is the NFW concentration parameter of the subhalo at accretion, $f(c)=\ln(1+c)-\frac{c}{1+c}$, $r_{\rm s,0}$ is the NFW scale radius at accretion, and $r_{\rm h}$ is the instantaneous half-mass radius of the subhalo.
We have verified that nearly all of the lowest mass subhalos considered in our HR simulations ($N_{\rm acc}\sim2\times10^4$ DM particles) verify these criteria.
Note however, that these criteria were generated from idealized DMO simulations that do not account for the disk potential present in our baryonic hosts, which more efficiently (and rapidly) disrupts subhalos that orbit close to the disk \citep[see e.g.][]{GK2017}.

\section{Differential radial distribution}\label{diff_radial_appendix}

\begin{figure}
	\includegraphics[width=\columnwidth]{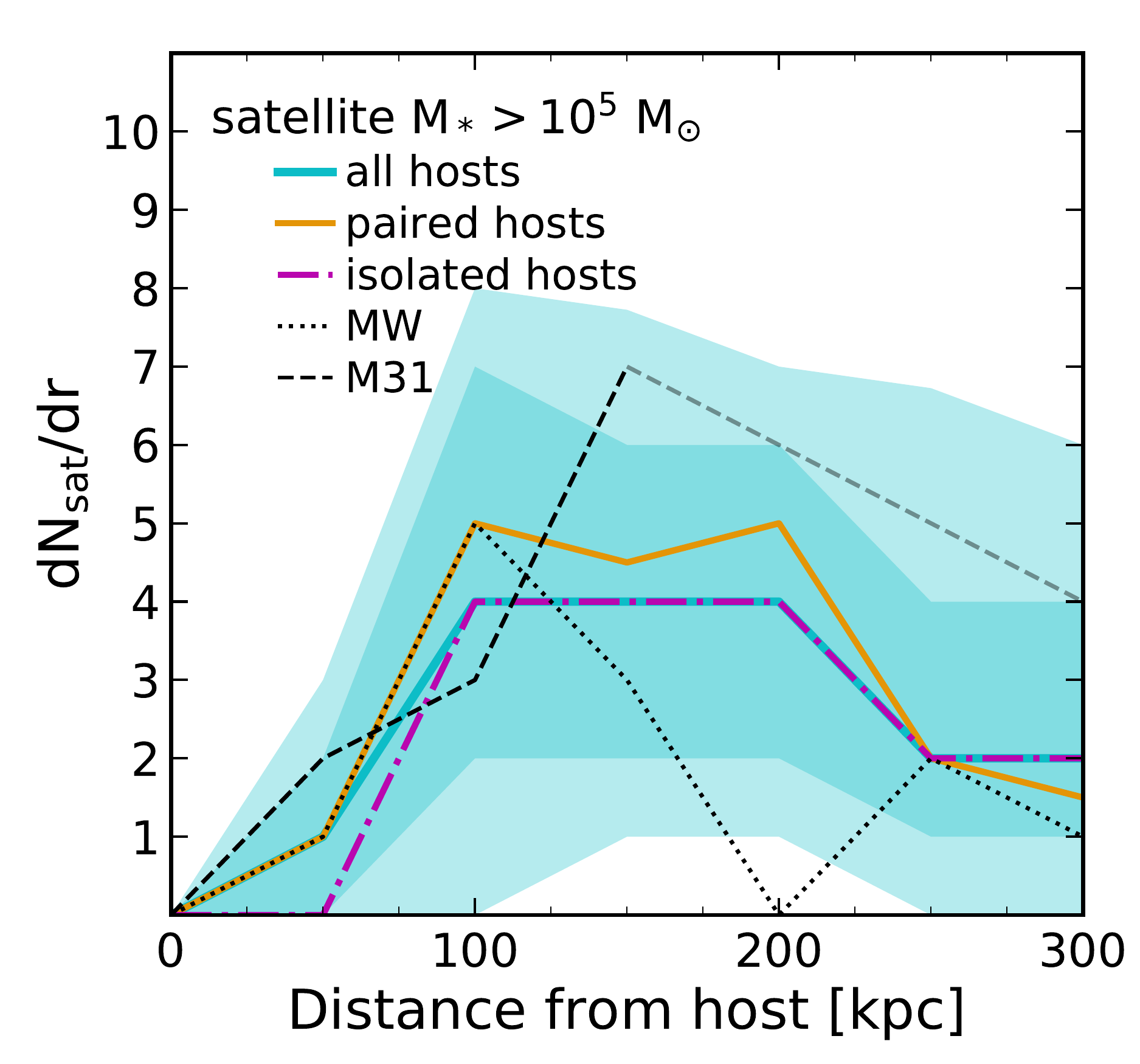}
	\vspace{-7 mm}
    \caption{Same as the top left panel of Figure~\ref{totalradialdist} for satellites with M$_* > 10^5$ M$_{\odot}$, except the radial distribution uses differential bins instead of cumulative.}
    \label{diffrad}
\end{figure}

We also examined the differentially-binned radial profiles of satellites around the hosts in our simulations. 
Figure~\ref{diffrad} shows these profiles for the simulations and MW/M31 observations, considering all satellites with M$_* > 10^5$ M$_{\odot}$. 
The simulation scatter (blue regions) encompasses both the MW and M31 profiles at the 68 per cent level out to about 150 kpc. 
Past this distance, the MW shows a known lack of satellites between $150-200$ kpc, but is otherwise consistent with the simulations at the 68 per cent level. 
Potential incompleteness in the MW's satellite population is explored further in Section~\ref{MW_inc}. 
M31 remains within the simulation scatter at the 95 per cent level from $150-300$ kpc. 
In general, the simulated differential distributions are a reasonable match to the Local Group.

\section{Correlation with host galaxy and halo mass}\label{host_mass_appendix}

\begin{figure}
	\includegraphics[width=\columnwidth]{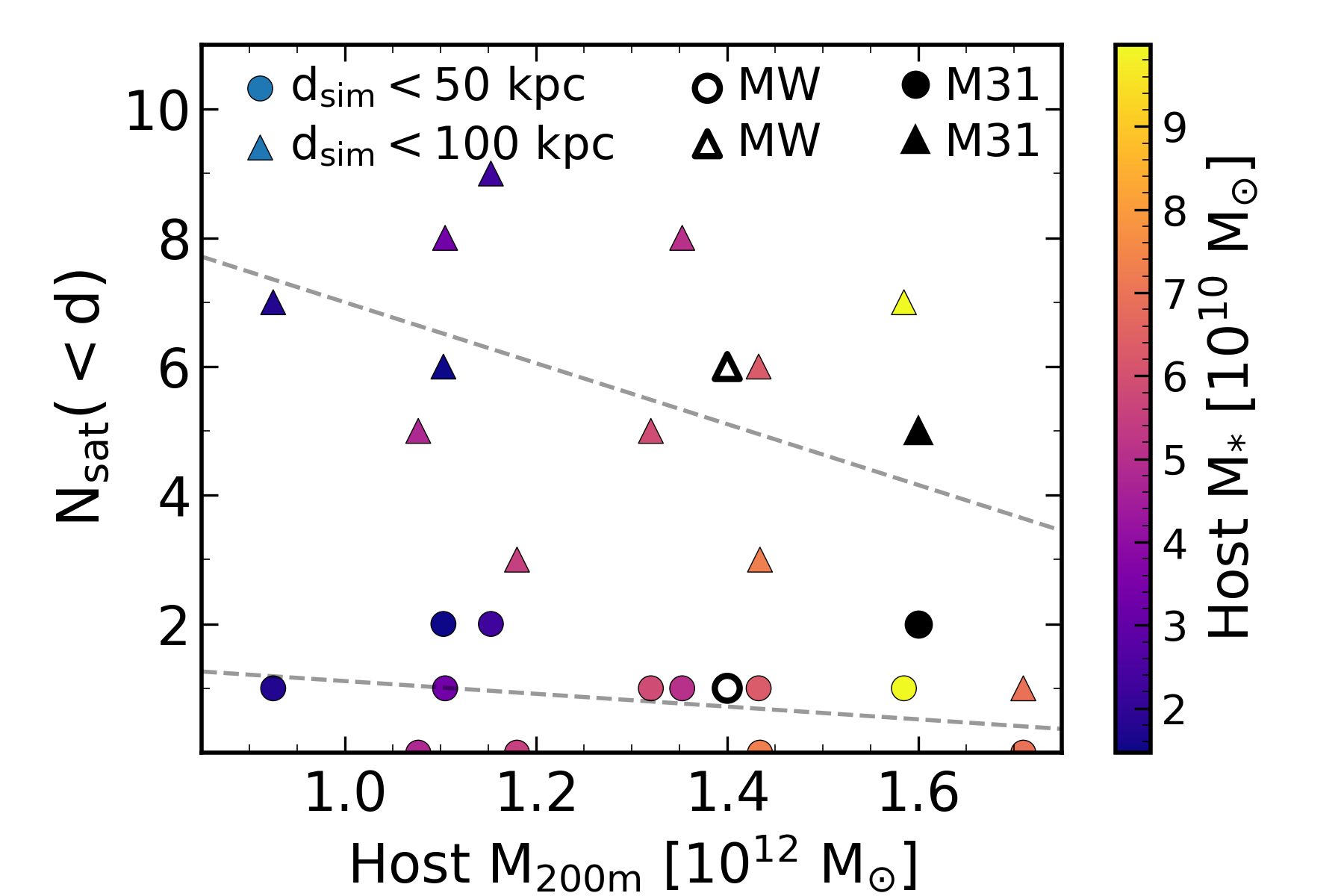}
	\vspace{-7 mm}
    \caption{Same as the right panel of Figure~\ref{nsat_v_host_mass} for satellites with M$_* > 10^5$ M$_{\odot}$, but the points have been colored by the mass of the host galaxy's baryonic disk. Scatter has been left off of the points for clarity.}
    \label{hostmasscolor}
\end{figure}

In Figure~\ref{hostmasscolor}, we repeat the exercise of section~\ref{host_mass}, but this time plotting the number of satellites at small distances as a function of host halo mass and color coding my host disk mass.
This illustrates that both host disk mass and halo mass simultaneously correlate with the number of satellites at small distances from the host.
Therefore, the main driver of the negative trends with host mass remains uncertain in our analysis.
However, other work that has systematically varied an analytical disk potential at fixed halo mass and found that the disk was the source of a reduction in DMO substructure close to the host when compared to a disk-less host halo \citep{Kelley2018}.
Although our results are not necessarily definitive on their own, we conclude that it is not unreasonable for the trends we see in N$_{\rm sat}$ as a function of host mass to originate from enhanced tidal destruction of satellites due to the host's baryonic disk.

%%%%%%%%%%%%%%%%%%%%%%%%%%%%%%%%%%%%%%%%%%%%%%%%%%

% Don't change these lines
\bsp	% typesetting comment
\label{lastpage}
\end{document}